\documentclass[review]{elsarticle}
\usepackage{setspace}
\usepackage{algorithm}
\usepackage{algorithmic}
\usepackage{graphicx}
\usepackage{epstopdf}
\usepackage{amsmath}
\usepackage{diagbox}
\usepackage{mathptmx}
\usepackage{caption}
\usepackage{subcaption}
\usepackage{booktabs}
\usepackage{amssymb}
\journal{e-Print}

\begin{document}
\doublespacing
\begin{frontmatter}
\title{A Novel Neuron Model of Visual Processor}

\author[LZU]{Jizhao Liu\corref{cor1}}
\ead{liujz@lzu.edu.cn}
\author[LJTU]{Jing Lian}
\author[UWM]{J C Sprott}
\author[LZU]{Yide Ma}
\cortext[cor1]{Corresponding author}
\address[LZU]{School of Information Science and Engineering, Lanzhou University, No.222, TianShui Road(south), Lanzhou, 730000, Gansu, China}
\address[LJTU]{School of Electronics and Information Engineering, Lanzhou Jiaotong University, No.88, West Anning Road, Lanzhou, 730070, Gansu, China}
\address[UWM]{Department of Physics, University of Wisconsin, Madison, WI 53706, USA.}

\begin{abstract}
Simulating and imitating the neuronal network of humans or mammals is a popular topic that has been explored for many years in the fields of pattern recognition and computer vision. Inspired by neuronal conduction characteristics in the primary visual cortex of cats, pulse-coupled neural networks (PCNNs) can exhibit synchronous oscillation behavior, which can process digital images without training. However, according to the study of single cells in the cat primary visual cortex, when a neuron is stimulated by an external periodic signal, the interspike-interval (ISI) distributions represent a multimodal distribution. This phenomenon cannot be explained by all PCNN models. By analyzing the working mechanism of the PCNN, we present a novel neuron model of the primary visual cortex consisting of a continuous-coupled neural network (CCNN). Our model inherited the threshold exponential decay and synchronous pulse oscillation property of the original PCNN model, and it can exhibit chaotic behavior consistent with the testing results of cat primary visual cortex neurons. Therefore, our CCNN model is closer to real visual neural networks. For image segmentation tasks, the algorithm based on CCNN model has better performance than the state-of-art of visual cortex neural network model. The strength of our approach is that it helps neurophysiologists further understand how the primary visual cortex works and can be used to quantitatively predict the temporal-spatial behavior of real neural networks. CCNN may also inspire engineers to create brain-inspired deep learning networks for artificial intelligence purposes.
\end{abstract}


\end{frontmatter}

\section{Introduction}
Creating brain-like machines is a long-standing goal in computer science. Neuromorphic computing efforts date back to the 1980s, when transistors were used to mimic the functionality of biological neurons and synapses. In the early 2000s, such research efforts (essentially spike-driven computations) facilitated the emergence of large-scale neuromorphic chips. Today, spiking neural networks (SNNs) are being actively explored using various algorithms. An SNN uses the time (pulse) of the signal to process information. Recently, research on SNNs has attracted the attention of many researchers from both hardware and algorithm aspects. Kaushik et al. stated that "the field of neuromorphic computing is a synergistic effort equally weighted across both hardware and algorithmic domains to enable spike-based artificial intelligence. "\cite{roy2019towards}
\par
According to Maass, neural networks can be categorized into three different generations\cite{maass1997networks}. The first generation is based on using McCulloch-Pitts neurons as computational units, which perform threshold operations and output Boolean results. Hopfield networks and Boltzmann machines are based on these neural network models. The second generation is based on computational units that output continuous values. By applying a sigmoid unit or a rectified linear unit (ReLU) as an "activation function", these networks are able to evaluate a continuous set of output values and address complex and deep tasks. The current deep learning networks, which have multiple hidden layers between their input and output, are all based on such second-generation neurons. The third generation of networks primarily use spiking neurons of the ‘integrate-and-fire’ type that exchange information via spikes. Scientists currently believe that mammals use spikes to process information.
\par
A pulse-coupled neural network (PCNN) is a third-generation artificial neural network proposed in the 1990s\cite{yang2019overview}. Inspired by the primary visual cortex of mammals such as cats, a PCNN can exhibit synchronous oscillation behavior and process digital images without training. In the past few decades, a large number of studies have proven that PCNN models achieve better performances in image segmentation\cite{lian2019image}, edge detection\cite{xie2016pcnn}, noise reduction\cite{shen2017hybrid}, fusion\cite{wang2015novel}, enhancement\cite{lian2019overview} and quantization\cite{yang2019study}.
\par
There are two main paths for studying PCNNs. The first path involves studying the parameter setting problem\cite{deng2019pcnn}, because a PCNN is a multiparameter iteration network, and the network’s performance depends on settings such as the iteration numbers, the decay time constants and linking weight coefficients. Broussard was the first to attempt to set parameters adaptively by using the gradient descent method\cite{broussard1997physiologically}. Yin et al. utilized the edge gradient to determine the linking coefficient $\beta $ of the PCNN, but other parameters were constants\cite{yin2017novel}. Li et al. proposed a novel autowave PCNN to solve the shortest path problem, adaptively setting the dynamic threshold according to the current network state\cite{li2013self}. Gao et al. proposed a modified PCNN mode by establishing a new neural threshold and a varying linking coefficient value\cite{gao2013automatic}. Lian et al. utilized a parameter-adaptive PCNN model to construct an automatic segmentation algorithm\cite{lian2017automatic,lian2017automatic1}.
\par
The second path proposes derivative models based on PCNN\cite{yang2019overview}. For example, Kinser simplified the PCNN model to the ICM model in 1996 and demonstrated that ICM was useful for target recognition\cite{kinser1999foveation}. In 2009, Zhan et al. proposed an SCM model by modifying the internal activity and firing condition of a PCNN. They showed that an SCM has lower computational complexity and achieves very high accuracy rates for image retrieval tasks compared with other common methods\cite{zhan2009new}. Two years later, Chen et al. simplified the SCM model to the SPCNN model by replacing the traditional firing condition\cite{chen2011new}. Deng, et al. found that the firing time is not integer-based and proposed a quasi-continuous PCNN model in 2012. In recent years, heterogeneous PCNN models have been proposed because researchers realized that the models should be more consistent with real biological nervous systems\cite{yang2018heterogeneous}.
\par
The PCNN and its derivative models have achieved great success on image processing tasks because they simulate the transmission characteristics of neurons in the cerebral cortex visual zone. Ironically, these models do not fully mimic all the characteristics of visual cortical neurons. Siegel used an external periodic signal to stimulate a cat's receptive fields and then recorded the signal of the primary visual cortex of the cat\cite{siegel1990non}. This experiment revealed that the interspike-interval (ISI) distributions represent a multimodal distribution, which all PCNNs cannot exhibit.
\par
In recent years, Bittner el al. found that synaptic plasticity relies on the relative timing of pre- and post-synaptic spikes\cite{bittner2017behavioral}. These features map to the synaptic weights of SNNs, including PCNN\cite{osswald2017spiking}. Therefore, it is very important to design a neuron model that generates more accurate predictions of stimulus responses. This motivated us to further study neuron models of the primary visual cortex. Using our method, not only can we predict the responses of neurons, but it may also inspire more efficient SNNs for artificial intelligence purposes.
\par
In this work, we present a novel neuron model of the primary visual cortex by analyzing the working mechanism of the PCNN, i.e., the continuous-coupled neural network (CCNN). To model the CCNN, we follow the minimal criteria for a sensory encoding model proposed by Yamins and DiCarlo, which are stimulus-computability, mappability, and predictivity\cite{yamins2016using}. Our model inherits the threshold exponential decay and synchronous pulse oscillation property from the original PCNN model, and its ISI distributions are closer to those of real neurons when stimulated by periodic signals. Thus, the CCNN model more closely approaches real visual neural networks and may benefit the next generation of deep learning networks.
\par
The rest of this paper is organized as follows. In Section II, we survey the relationship between the PCNN model and the results of neurophysiological experiments and analyze the output spike train data of the PCNN. Then, we analyzed the deficiencies and limitations of designing such SNNs for artificial intelligence purposes. In Section III, we introduce the CCNN model and analyze its dynamic behavior. In Section IV, we investigate the spiking feature of CCNN, demonstrate why the CCNN model is closer to the results of neurophysiological experiments and analyze how the CCNN may further benefit artificial intelligence. In Section V, we propose and evaluate an image segmentation algorithm; the comparative results indicate that our method achieves better performance than the image segmentation algorithm without training. Finally, Section VI provides conclusions and suggests future work.

\section{Pulse-Coupled Neural Network}
\subsection{The biological feature of neurons}
1. Neurons do not create an action potential (AP) immediately; instead, they undergo a refractory period\cite{novak2004refractory}. The absolute refractory period (ARP) refers to the period during which a neuron creates two consecutive APs. The ARP is caused by the inactivation of the Na+ channels that originally opened to depolarize the membrane. These channels then remain inactivated until the membrane hyperpolarizes. The channels then close, de-inactivate, and regain their ability to open in response to a stimulus. The relative refractory period immediately follows the ARP. As the voltage-gated potassium channels open to terminate the action potential by repolarizing the membrane, the potassium conductance of the membrane increases dramatically. The K+ ions moving out of the cell move the membrane potential closer to the equilibrium potential for potassium, which causes a brief hyperpolarization of the membrane—that is, the membrane potential becomes transiently more negative than its normal resting potential. Until the potassium conductance returns to the resting value, a larger stimulus will be required to reach the initiation threshold to initiate a second depolarization. This period is called the relative refractory period (RRP).
\par
2. In the primary visual cortex, coherent oscillations exist, which means that neurons create action potentials not only due to external stimuli but also from stimuli by other neurons. This process is called stimulus-induced movement. The original experiments were performed by Eckhorn et al., who recorded single and multiple spikes as well as local field potentials simultaneously from several locations in the primary visual cortex (A17 and A18) of a cat. They reported that "Coherent stimulus-evoked resonances were found at distant cortical positions when at least one of the primary coding properties was similar. Coherence was found 1) within a vertical cortex column, 2) between neighboring hypercolumns, and 3) between two different cortical areas"\cite{eckhorn1988coherent}.
\par
\subsection{The genesis of PCNN}
Based on the above biological features, a pulse-coupled neural network (PCNN) was proposed\cite{lindblad2005image}. The basic model’s main parts are couple linking, feeding input, modulation product, dynamic threshold, and spiking output. The couple-linking part represents the spiking signal received from other neurons. The feeding input part is the signal from external sources through the receptive fields, which can be pulses, analog time-varying signals, constants, or any combination. The received input signals are modulated through the modulation product part. The dynamic threshold part is used to map the refractory period of neurons. Spiking the output part creates an action potential signal when the signal satisfies the burst conditions. The model is shown in Fig. \ref{fig:PCNN}.
\begin{figure}[!htb]
	\centering
	\includegraphics[width=0.6\linewidth]{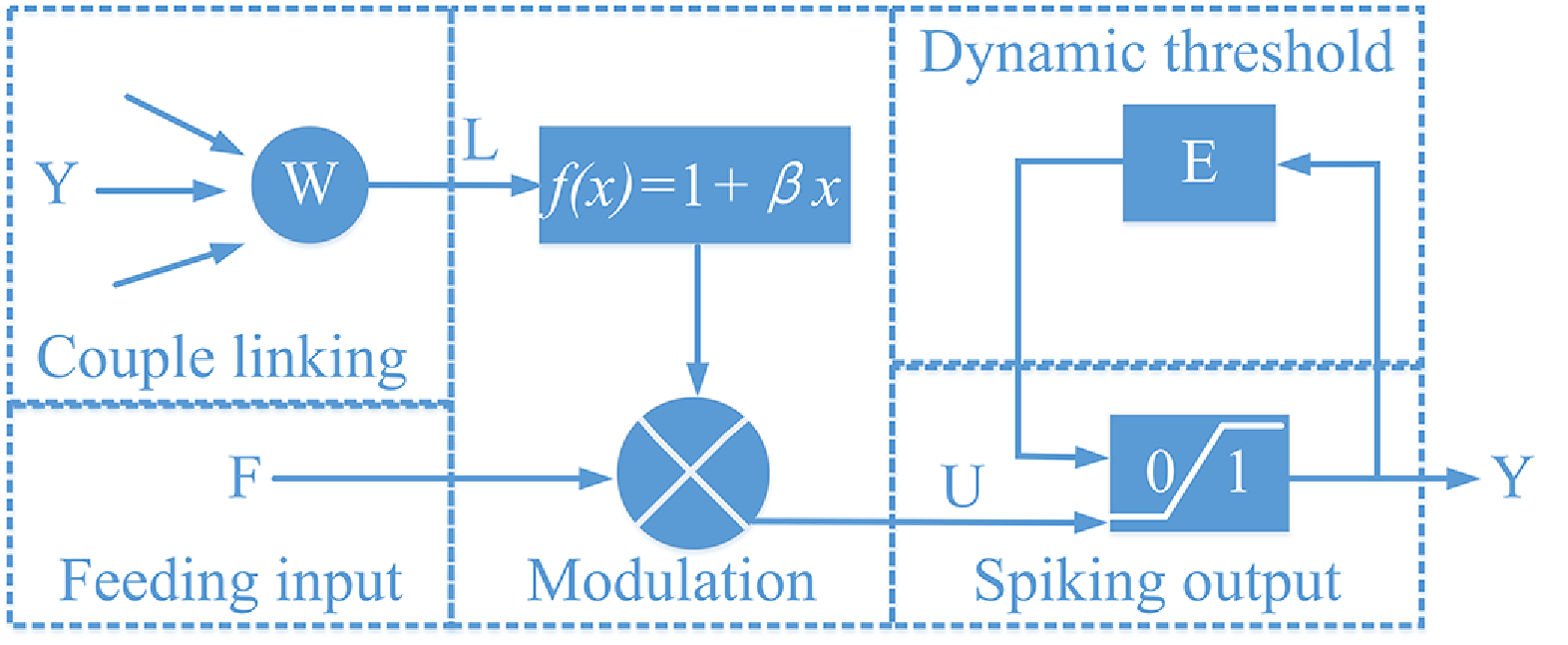}
	\setcounter{figure}{0}
	\caption{Pulse-coupled neuron model}
	\label{fig:PCNN}
\end{figure}
\par 
The following equation further describes the PCNN model:
\begin{equation} \label{Eq:PCNN}
	\begin{split}
		\begin{aligned}
			F_{ij}[n]&=e^{-\alpha_{f}}F_{ij}[n-1]+V_{F}M_{ijkl}Y_{kl}[n-1]+S_{ij}\\
			L_{ij}[n]&=e^{-\alpha_{l}}L_{ij}[n-1]+V_{L}W_{ijkl}Y_{kl}[n-1]\\
			U_{ij}[n]&=F_{ij}[n](1+\beta L_{ij}[n])\\
			Y_{ij}[n]&=\left\{
			\begin{aligned}
				1, & \qquad if U_{ij}[n]>E_{ij}[n]\\
				0, & \qquad otherwise \\
			\end{aligned}
			\right. \\
			E_{ij}[n]&=e^{-\alpha_{e}}E_{ij}[n-1]+V_{E}Y_{ij}[n]
		\end{aligned}
	\end{split}, 
\end{equation}
where the five main parts are couple linking $ L_{ij}[n] $, feeding input $ F_{ij}[n] $, modulation product $ U_{ij}[n] $, dynamic threshold $ E_{ij}[n] $ and spiking output $ Y_{ij}[n] $. $ S_{ij} $ is the external feeding input stimulus received by the receptive fields. $ \alpha_{f} $, $ \alpha_{l} $ and $ \alpha_{e} $ denote exponential decay factors that record previous input states. $ V_{F} $ and $ V_{L} $ are weighting factors modulating the action potentials of surrounding neurons. Additionally, $ M_{ijkl} $ and $ W_{ijkl} $ denote the feeding and linking synaptic weights, respectively, and $ \beta $ denotes the linking strength, which directly determines $ L_{ij}[n] $ in the modulation product $ U_{ij}[n] $.
\par
From the above equations, the exponential decay factors ingeniously imitate the refractory period because if a neuron creates an action potential or a spike, the dynamic threshold will increase to a large value, but it decays as time passes; therefore, it will be easier to burst after a the refractory period. The couple linking part imitates coherent oscillation behavior if the right values are chosen.
\par 
Since PCNN and its derivative models can simulate the transmission characteristics of neurons in the cerebral cortex visual zone, they have achieved great success on some unsupervised image processing tasks.
\subsection{The deficiencies and the limitations of PCNN}
Action potentials are propagated along the axons of a neuron to reach the nerve terminals, where they can trigger the release of chemical messengers that affect other neurons. Therefore, spikes are an important way in which neurons in the brain carry information\cite{dettner2016temporal}. The interspike interval (ISI) is the time between successive action potentials (also known as spikes) of a neuron. Thus, ISI distributions are believed to be an effective tool for revealing the mechanism of brain code information\cite{aljadeff2016analysis}.
\par
\begin{figure}[!htb]
	\centering
	\includegraphics[width=0.3\linewidth]{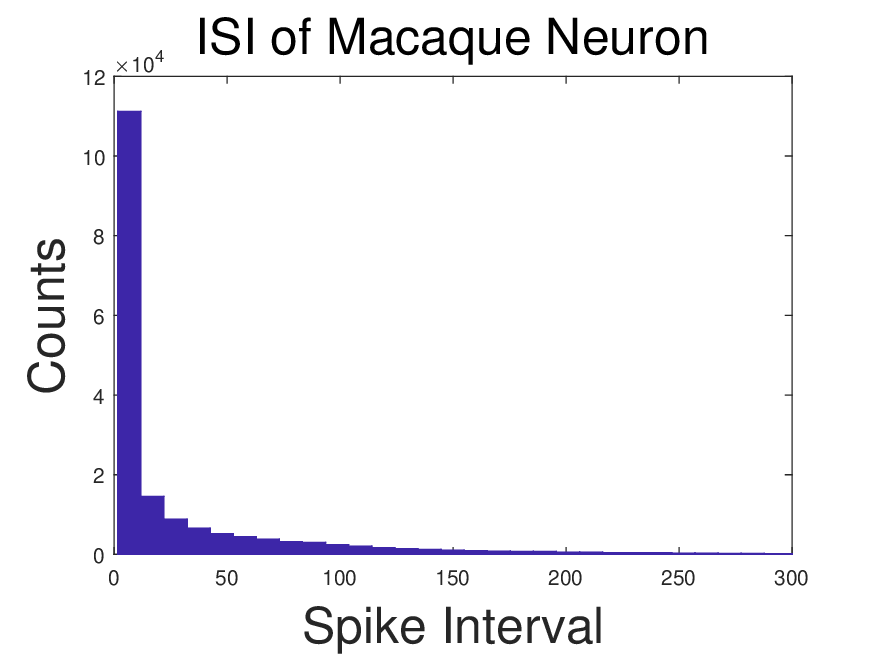}
	\caption{Interspike interval of real neurons under natural image external stimulus}
	\label{fig:RealISI}
\end{figure}

 
Based on the data which represent the responses of V1 populations of neurons to natural image sequences \cite{ringach2016spatial}, the spike train of a single neuron of the primary visual cortex (V1 area) of macaques can be calculated, which is shown as Fig. \ref{fig:RealISI}.\footnote{The data used here were collected in the laboratory of Dario Ringach at UCLA and downloaded from the CRCNS website. https://crcns.org/data-sets/vc/pvc-1/conditions}. Fig. \ref{fig:RealISI} indicate that the ISI distributions of a single neuron have many different periods when stimulated by natural images.
\par
More detailed experiments were performed by Siegel, who used a hand-held projector to stimulate the cells of receptive fields with a high-speed square wave signal and collected the response spike train of a single neuron of the primary visual cortex of a cat (area 17). He found many patterns in the return maps of ISI distributions, see Fig. 5 in ref.\cite{siegel1990non}.
\par
\begin{figure}[htbp]
	\begin{minipage}[t]{0.3\linewidth}
		\centering
		\includegraphics[width=1\linewidth]{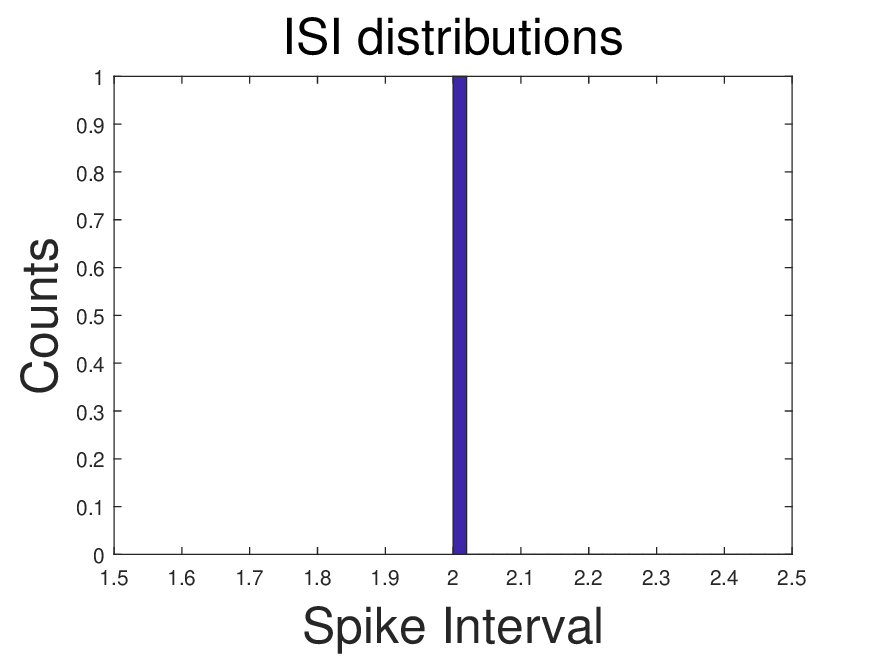}\\
		\subcaption{}
		\vspace{0.02cm}
	\end{minipage}
	\begin{minipage}[t]{0.3\linewidth}
		\centering
		\includegraphics[width=1\linewidth]{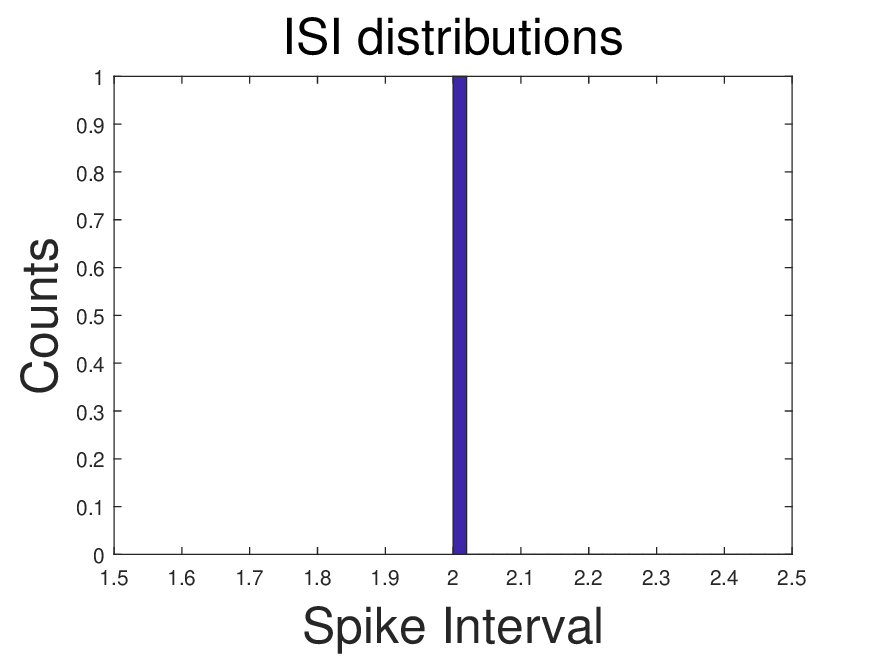}\\
		\subcaption{}
		\vspace{0.02cm}
	\end{minipage}%
	\centering
	\caption{Interspike interval of PCNN neurons under different stimuli. (a) $ S=0.5(1+sin(t)) $; (b) $ S= $Logistic Map }
	\label{fig:PCNNISI}
\end{figure}

However, the PCNN can output only a fixed-period response signal, which leads the ISI distributions of the PCNN to manifest as a single bar. To illustrate further, we use sine functions (periodically driven signals) and logistic maps (chaotically driven signals) as external stimulus signals $ S_{ij} $ and calculate the ISI distribution of the response spike train of the PCNN, as shown in Fig.\ref{fig:PCNNISI}. Therefore, PCNN cannot mimic the dynamic behavior of real neurons.
\par
On the other hand, the general idea for using SNNs in artificial intelligence is to use the spike intervals as a type of error that can be backpropagated by algorithms\cite{bohte2002error}. The PCNN only outputs a spiking train with a fixed period. Therefore, it would seem impossible to utilize PCNN to design SNNs for artificial intelligence, despite PCNN having achieved great success on image processing tasks.
\par
Based on the above analysis, a PCNN cannot predict the temporal-spatial behavior of real neural networks, and it cannot benefit artificial intelligence as other SNNs can. This limitation prevented further research and application into PCNNs.
\section{Continuous-Coupled Neural Network}
These deficiencies and limitations of the PCNN model motivated us to further explore models of the primary visual cortex. We followed Yamins' criteria to design the CCNN. As described in\cite{yamins2016using}, these are:
\par
Stimulus computability: The model should accept arbitrary stimuli within the general stimulus domain of interest;
\par
Mappability: The components of the model should correspond to experimentally definable components of the neural system; and
\par
Predictivity: The units of the model should provide detailed predictions of stimulus-by-stimulus responses for arbitrarily chosen neurons in each mapped area.
\par
\subsection{CCNN Models}
\begin{figure}[!htb]
	\centering
	\includegraphics[width=0.6\linewidth]{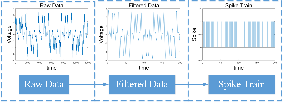}
	\caption{All-or-none coding with action potentials}
	\label{fig:flow chart of coding AP}
\end{figure}

In neurophysiology experiments, nerve action potentials in response to changing stimuli are monitored and recorded by an electrode. Then, bandpass filtering is utilized to separate the relatively high frequency components in the action potential from low frequency noise, which stems from measurement equipment and the environment. Finally, the action potential is coded to the spike train by the all-or-none law: if the voltage exceeds the threshold potential, the nerve fiber will generate a complete response; otherwise, it provides no response. A flowchart of the coding action potential is shown in Fig.\ref{fig:flow chart of coding AP}.
\par
\begin{figure}[!htb]
	\centering
	\includegraphics[width=0.6\linewidth]{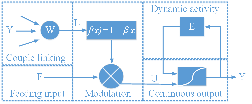}
	\caption{Continuous coupled neuron model}
	\label{fig:CCNN}
\end{figure}

Therefore, it is reasonable to modify the spiking part of the PCNN (Fig. \ref{fig:PCNN}) to a continuous output part (Fig. \ref{fig:CCNN}), to create the CCNN. When studying the spiking features of CCNN, one can generate the spiking train data by using an appropriate step function as the filter. The CCNN model is shown in Fig. \ref{fig:CCNN}.
\par
The basic model’s main parts are couple linking, feeding input, modulation product, dynamic threshold, and continuous output. The couple linking part receives output signals from other neighboring neurons. The feeding input part consists of signals from external sources through the receptive fields. The modulation product part calculates all the received input signals. The dynamic activity part is used to map the refractory period of neurons. The continuous output part generates the response signal of the CCNN.
\par
The following equation further describes the CCNN model:
\begin{equation} \label{Eq:CCNN}
	\begin{split}
		\begin{aligned}
			F_{ij}[n]&=e^{-\alpha_{f}}F_{ij}[n-1]+V_{F}M_{ijkl}Y_{kl}[n-1]+S_{ij}\\
			L_{ij}[n]&=e^{-\alpha_{l}}L_{ij}[n-1]+V_{L}W_{ijkl}Y_{kl}[n-1]\\
			U_{ij}[n]&=F_{ij}[n](1+\beta L_{ij}[n])\\
			Y_{ij}[n]&=f(U_{ij}[n]-E_{ij}[n])\\
			E_{ij}[n]&=e^{-\alpha_{e}}E_{ij}[n-1]+V_{E}Y_{ij}[n-1]
		\end{aligned}
	\end{split}.
\end{equation}
\par
\subsection{The nonlinear function of continues output}
\begin{figure}[htbp]
	\centering
	\begin{minipage}[t]{0.24\linewidth}
		\centering
		\includegraphics[width=1\linewidth]{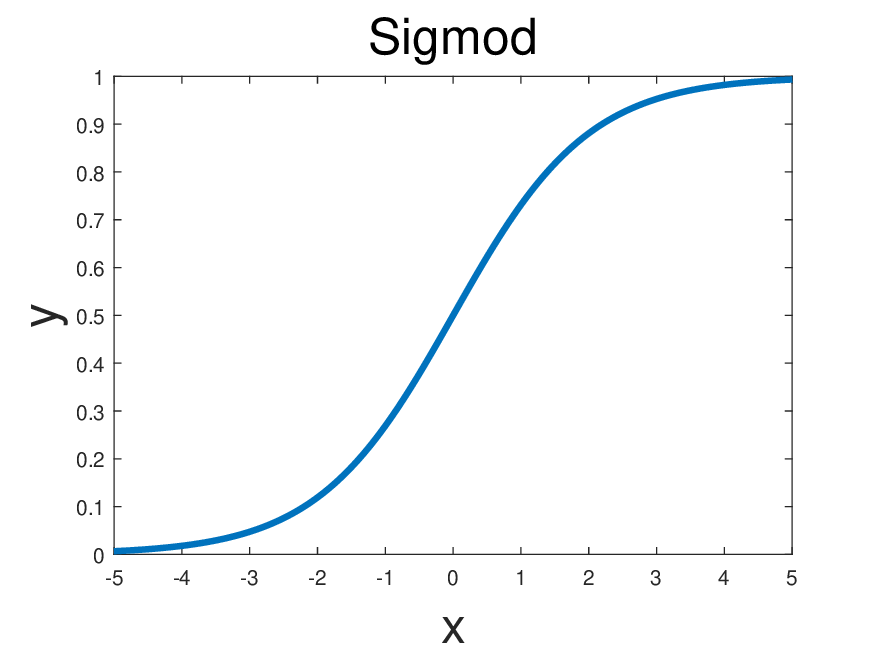}
		\vspace{0.02cm}
		\subcaption{}
	\end{minipage}
	\begin{minipage}[t]{0.24\linewidth}
		\centering
		\includegraphics[width=1\linewidth]{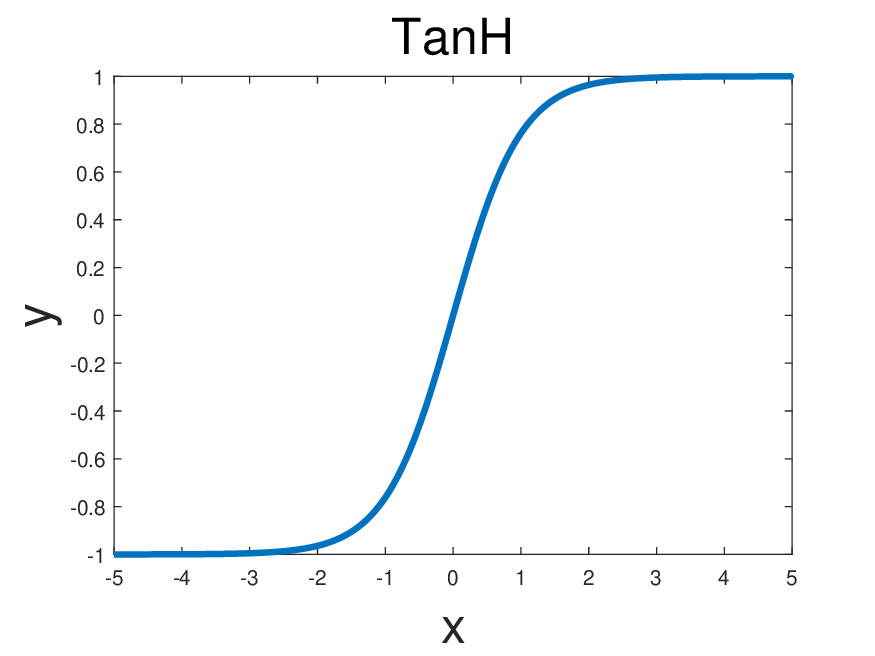}
		\vspace{0.02cm}
		\subcaption{}
	\end{minipage}
	\begin{minipage}[t]{0.24\linewidth}
		\centering
		\includegraphics[width=1\linewidth]{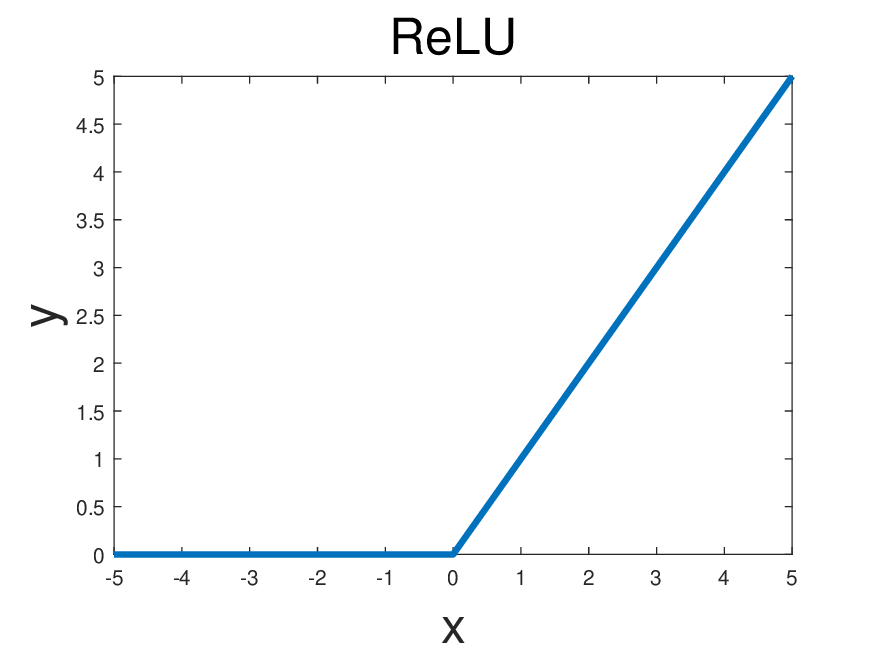}
		\vspace{0.02cm}
		\subcaption{}
	\end{minipage}
	\begin{minipage}[t]{0.24\linewidth}
		\centering
		\includegraphics[width=1\linewidth]{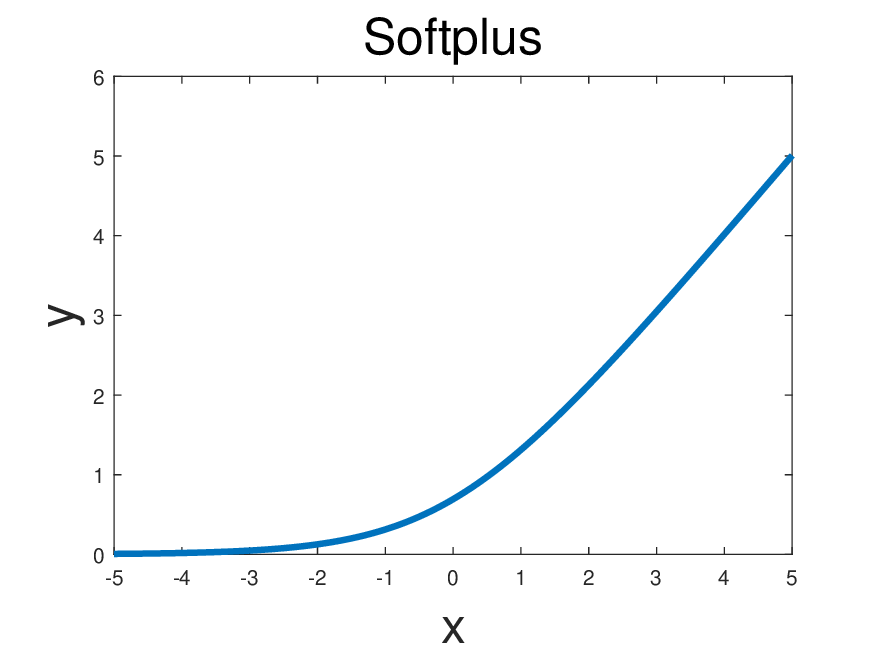}
		\vspace{0.02cm}
		\subcaption{}
	\end{minipage}%
	\centering
	\caption{Function graph of the nonlinear functions: (a) Sigmoid; (b) TanH; (c) ReLU; (d) Softplus}
	\label{fig:Function graph of nonlinear functions}
\end{figure}

As it is well known, the dynamic characteristics of a dynamic system vary with the nonlinear function. In order to design a system which has similar dynamic characteristics to real neurons, we use Sigmoid, TanH, ReLU and Softplus as the nonlinear function to test the dynamic behavior and ISI distributions. The function graph of the nonlinear functions are shown in Fig.\ref{fig:Function graph of nonlinear functions}. When the system is stimulated by DC signals, the results are shown in Tab.\ref{tab:dynamic behavior under DC signals}. When the system is stimulated by periodic signals, the results are shown in Tab.\ref{tab:dynamic behavior under periodic signals}. 
\begin{table}[htbp] 
	\centering
	{\begin{tabular}{lccl}
			\toprule
			Nonlinear Function & Equation & Dynamic Behavior\\
			\midrule
			Sigmoid  & $ f(x)=\frac{1}{1+e^{-x}} $ & Periodic Behavior\\
			TanH     & $ f(x)=\frac{e^{x}-e^{-x}}{e^{x}+e^{-x}} $ & Periodic Behavior\\
			ReLU     & $ f(x)=max(0,x) $     & Periodic Behavior\\
			Softplus & $ f(x)=log(1+e^{x}) $ & Chaotic Behavior\\
			\bottomrule
	\end{tabular}}
	\caption{Dynamic behavior under DC signals}
	\label{tab:dynamic behavior under DC signals}
\end{table}
\par
\begin{table}[htbp] 
	\centering
	{\begin{tabular}{lccl}
			\toprule
			Nonlinear Function & Equation & Chaotic Behavior\\
			\midrule
			Sigmoid  & $ f(x)=\frac{1}{1+e^{-x}} $ & Chaotic Behavior\\
			TanH     & $ f(x)=\frac{e^{x}-e^{-x}}{e^{x}+e^{-x}} $ & Periodic Behavior\\
			ReLU     & $ f(x)=max(0,x) $     & Periodic Behavior\\
			Softplus & $ f(x)=log(1+e^{x}) $ & Chaotic Behavior\\
			\bottomrule
	\end{tabular}}
	\caption{Dynamic behavior under periodic signals}
	\label{tab:dynamic behavior under periodic signals}
\end{table}
\par
According to Hodgkin and Siegel, real neurons will exhibit periodic behavior under DC drive signal, and they will exhibit chaotic behavior under periodic drive signal. Therefore, using Sigmoid as the nonlinear function of CCNN model is a reasonable choice. The CCNN model is described by:
\begin{equation} \label{Eq:CCNN1}
	\begin{split}
		\begin{aligned}
			F_{ij}[n]&=e^{-\alpha_{f}}F_{ij}[n-1]+V_{F}M_{ijkl}Y_{kl}[n-1]+S_{ij}\\
			L_{ij}[n]&=e^{-\alpha_{l}}L_{ij}[n-1]+V_{L}W_{ijkl}Y_{kl}[n-1]\\
			U_{ij}[n]&=F_{ij}[n](1+\beta L_{ij}[n])\\
			Y_{ij}[n]&=1/(1+e^{-(U_{ij}[n]-E_{ij}[n])})\\
			E_{ij}[n]&=e^{-\alpha_{e}}E_{ij}[n-1]+V_{E}Y_{ij}[n-1]
		\end{aligned}
	\end{split}.
\end{equation}
\par
Later, we will illustrate that the CCNN model can exhibit both periodic and chaotic behavior, causing the model to generate numerous spike train sequences with a threshold filter. This property empowers the CCNN model to achieve better performances on image processing applications and may further benefit the use of spiking neural networks for artificial intelligence purposes. The various spike train patterns are also more similar to those of real visual networks.
\subsection{Dynamic behavior of CCNN: Noncoupling with neighborhood neurons}
\subsubsection{Constant stimulus}
The single neuron model can be simplified from Eq. (\ref{Eq:CCNN}) by setting $ M_{ijkl}=0 $, $ W_{ijkl}=0 $ and $ \beta=0 $. The equations of the single neuron model are given by:
\begin{equation} \label{Eq:singleCCNN}
	\begin{split}
		\begin{aligned}
			F[n]&=e^{-\alpha_{f}}F[n-1]+S\\
			U[n]&=F[n]\\
			Y[n]&=1/(1+e^{-(U[n]-E[n])})\\
			E[n]&=e^{-\alpha_{e}}E[n-1]+V_{E}Y[n-1]
		\end{aligned}
	\end{split}.
\end{equation}
\par
When $ S=0 $, the nerve neuron has no input signal, and Eq.\ref{Eq:CCNN} can be simplified to:
\begin{equation} \label{Eq:singleCCNN1}
	\begin{split}
		\begin{aligned}
			F[n]&=0\\
			U[n]&=F[n]=0\\
			Y[n]&=1/(1+e^{E[n]})\\
			E[n]&=e^{-\alpha_{e}}E[n-1]+V_{E}Y[n-1]
		\end{aligned}
	\end{split}.
\end{equation}
Substituting the fourth equation into the fifth equation, we obtained the recursive formula of $ E_{n} $, which is:
\begin{equation} \label{Eq:singleCCNN2}
	\begin{split}
		\begin{aligned}
			E_{n}= e^{-\alpha_{e}}E[n-1]+V_{E}/(1+e^{E[n-1]})
		\end{aligned}
	\end{split}.
\end{equation}
The fixed point of Eq. (\ref{Eq:singleCCNN2}) can be calculated by: 
\begin{equation} \label{Eq:singleCCNN3}
	\begin{split}
		\begin{aligned}
			E=e^{-\alpha_{e}}E+V_{E}/(1+e^{E})
		\end{aligned}
	\end{split};
\end{equation}
therefore, the fixed points are as follows:
\begin{equation} \label{Eq:singleCCNN4}
	\begin{split}
		\begin{aligned}
			E(1+e^{E})=V_{E}/(1-e^{-\alpha_{e}})
		\end{aligned}
	\end{split}.
\end{equation}
Since $ f(E)=E(1+e^{E}) $ is a monotonically increasing function, when $ E \rightarrow -\infty $, $ f(E)\rightarrow 0 $. Thus, when $ V_{E}/(1-e^{-\alpha_{e}})>0 $, the CCNN model has only one fixed point; otherwise, it has no fixed point. In fact, $ V_{E} $ is the reset voltage, which is always positive, and the CCNN has only one fixed point if and only if $ \alpha_{e}>0 $.
\par
Next, we investigate whether this fixed point is stable. We have the following theorem.
\par
\textbf{Theorem:} The necessary condition for the fixed points or period-1 orbit to be stable is that the absolute value of the first derivative of Eq. (\ref{Eq:singleCCNN3}) must be less than $ 1 $.
\par
\textbf{Proof:} For a nonlinear function $ x=f(x) $, assume that a slight disturbance occurs at the fixed point $ x^{*} $. The equation of the fixed point is: 
\begin{equation} \label{Eq:fixedPoint1}
	\begin{split}
		\begin{aligned}
			x^{*}+\varepsilon_{n+1}=f(x^{*}+\varepsilon_{n}) 
		\end{aligned}
	\end{split},
\end{equation}
where $ \varepsilon_{n} $ is the distance before the disturbance and $ \varepsilon_{n+1} $ is the distance after the disturbance. We expand the right-hand side of Eq. (\ref{Eq:fixedPoint1}) and obtain the linear term of $ \varepsilon_{n} $ as follows:
\begin{equation} \label{Eq:fixedPoint2}
	\begin{split}
		\begin{aligned}
			x^{*}+\varepsilon_{n+1}=f(x)+\frac{\partial f(x)}{\partial x}|_{x=x^{*}}\varepsilon_{n}+\cdots
		\end{aligned}
	\end{split}.
\end{equation}
Substituting $ x^{*}=f(x^{*}) $ into Eq. (\ref{Eq:fixedPoint2}), we have: 
\begin{equation} \label{Eq:fixedPoint3}
	\begin{split}
		\begin{aligned}
			\dfrac{\varepsilon_{n+1}}{\varepsilon_{n}}=\frac{\partial f(x)}{\partial x}|_{x=x^{*}}\varepsilon_{n}
		\end{aligned}
	\end{split}.
\end{equation}
For a stable fixed point, $ |\varepsilon_{n+1}|<|\varepsilon_{n}| $. Therefore, the stable condition of the fixed point is: 
\begin{equation} \label{Eq:fixedPoint4}
	\begin{split}
		\begin{aligned}
			s=\frac{\partial f(x)}{\partial x}|_{x=x^{*}}\varepsilon_{n} \leq 1
		\end{aligned}
	\end{split},
\end{equation}
where $ s=1 $ is the stable boundary. For $ f'(x)=1 $, the dynamic system exhibits tangent bifurcation behavior, while for $ f'(x)=-1 $, the dynamic system exhibits period doubling bifurcation behavior.
\par
The derivative function of the fixed point is given by: 
\begin{equation} \label{Eq:fixedPoint5}
	\begin{split}
		\begin{aligned}
			\frac{\partial f(E)}{\partial E}=e^{-\alpha_{e}}-V_{E}*\frac{e^{E}}{(1+e^{E})^{2}}
		\end{aligned}
	\end{split}.
\end{equation}
When $ |\frac{\partial f(E)}{\partial E}|\leq 1 $, the fixed points are stable. To solve the critical value of Eq. (\ref{Eq:fixedPoint5}), we can use the variable substitution method. Assuming $ y=e^{E} $, Eq. (\ref{Eq:fixedPoint5}) can be simplified to: 
\begin{equation} \label{Eq:fixedPoint9}
	\begin{split}
		\begin{aligned}
			\frac{y}{(1+y)^{2}}=\frac{1\pm e^{-\alpha_{e}}}{V_{E}}
		\end{aligned}
	\end{split}.
\end{equation}
Since $ y=e^{E}\geq 0 $, when $ f'(y)=\frac{1-2y}{(1+y)^{2}}=0 $, $ y=0.5 $, $ E=ln0.5 $, $ f(y) $ reached its extreme point, $ f_{max}(y)=\frac{y}{(1+y)^{2}}=0.22 $. Therefore, if $ \frac{1\pm e^{-\alpha_{e}}}{V_{E}}\leq 0.22 $, $ ln(-0.22V_{E}-1)\leq\mp\alpha_{e}\leq ln(0.22V_{E}-1) $, the fixed point is stable; otherwise, the fixed point is unstable.
\par
When $ S=C, F[0]=0$, $ F[n]=e^{-\alpha_{f}}F[n-1]+C=C(1+e^{-\alpha_{e}}+e^{-2\alpha_{e}}+\cdots+e^{-(n-1)\alpha_{e}})=C(1-e^{-n\alpha_{e}})/(1-e^{\alpha_{e}})=C_{S}$. Eq. (\ref{Eq:singleCCNN1}) can be simplified to: 
\begin{equation} \label{Eq:singleCCNN5}
	\begin{split}
		\begin{aligned}
			Y[n]&=1/(1+e^{E[n]-C_{S}})\\
			E[n]&=e^{-\alpha_{e}}E[n-1]+V_{E}Y[n-1]
		\end{aligned}
	\end{split}.
\end{equation}
Substituting $ Y[n-1] $ into $ E[n] $, we obtain the recursive formula of $ E_{n} $, which is:
\begin{equation} \label{Eq:singleCCNN6}
	\begin{split}
		\begin{aligned}
			E_{n}= e^{-\alpha_{e}}E[n-1]+V_{E}/(1+e^{E[n-1]-C_{S}})
		\end{aligned}
	\end{split}.
\end{equation}
Therefore, the fixed points are: 
\begin{equation} \label{Eq:singleCCNN7}
	\begin{split}
		\begin{aligned}
			E(1+e^{E-C_{S}})=V_{E}/(1-e^{-\alpha_{e}})
		\end{aligned}
	\end{split}.
\end{equation}
The derivative function of the fixed point is given by:
\begin{equation} \label{Eq:fixedPoint6}
	\begin{split}
		\begin{aligned}
			\frac{\partial f(E)}{\partial E}=e^{-\alpha_{e}}-V_{E}*\frac{e^{E-C_{S}}}{(1+e^{E-C_{S}})^{2}}
		\end{aligned}
	\end{split}.
\end{equation}
When $ E=ln0.5+C_{S} $, $ \frac{e^{E-C_{S}}}{(1+e^{E-C_{S}})^{2}} $ reaches its extreme point of $ 0.22 $. If $ \frac{1\pm e^{-\alpha_{e}}}{V_{E}}\leq 0.22 $, $ ln(-0.22V_{E}-1)\leq\mp\alpha_{e}\leq ln(0.22V_{E}-1) $, the fixed point is stable; otherwise, the fixed point is unstable.
\par
Let $ E_{n}=h(n) $; then, the derivative function of Eq. (\ref{Eq:singleCCNN6}) can be given by: 
\begin{equation} \label{Eq:fixedPoint7}
	\begin{split}
		\begin{aligned}
			\frac{\partial f(E)}{\partial n}&=\frac{\partial f(E)}{\partial E}*\frac{\partial E}{\partial n}=\frac{\partial f(E)}{\partial E}*\frac{\partial h(n)}{\partial n}\\
			&=(e^{-\alpha_{e}}-V_{E}*\frac{e^{E-C_{S}}}{(1+e^{E-C_{S}})^{2}})*\frac{\partial h(n)}{\partial n}
		\end{aligned}
	\end{split}.
\end{equation}
\par
If $ f^{T}(E)=E $, then $ T $ is the period of $ E $. Since $ e^{E-C_{S}}>0 $, $ g(C_{S})=e^{-\alpha_{e}}-V_{E}*\frac{e^{E-C_{S}}}{(1+e^{E-C_{S}})^{2}} $ is a monotonically increasing function. Therefore, for a specific period of fixed points of $ E $, a larger $ C_{S} $ will lead to a larger $ {\partial f(E)}/{\partial n} $; therefore, the fixed period-1 orbit will return to its starting point faster. In other words, increasing $ C_{S} $ decreases the periods of $ E $ and $ Y $. This phenomenon can be verified through an experiment, as shown in Fig.\ref{fig:CCNNsti}.
\par
\subsubsection{Periodic stimulus}
\begin{figure}[htbp]
	\centering
	\begin{minipage}[t]{0.3\linewidth}
		\centering
		\includegraphics[width=1\linewidth]{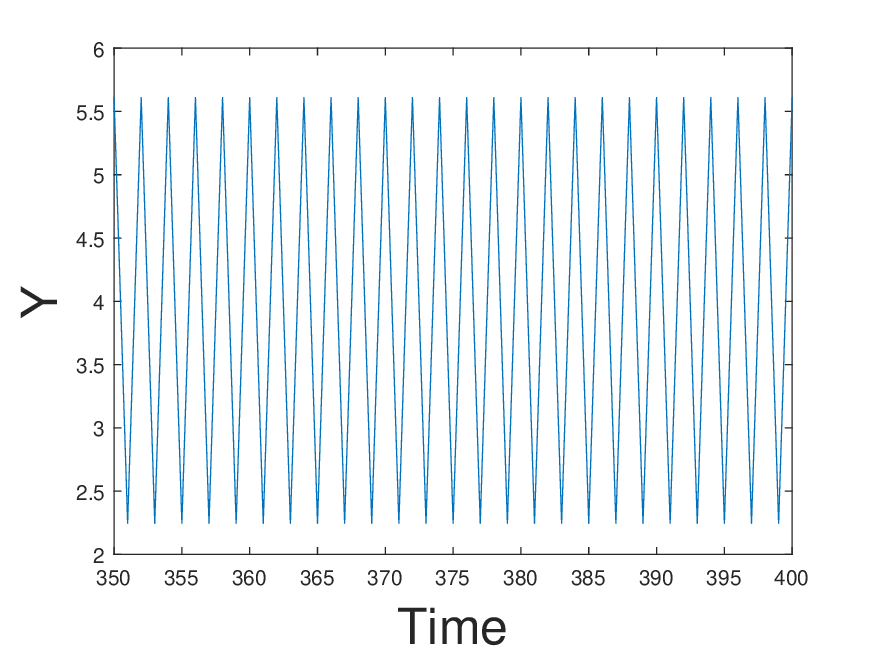}\\
		\vspace{0.02cm}
		\subcaption{}
		\centering
		\includegraphics[width=1\linewidth]{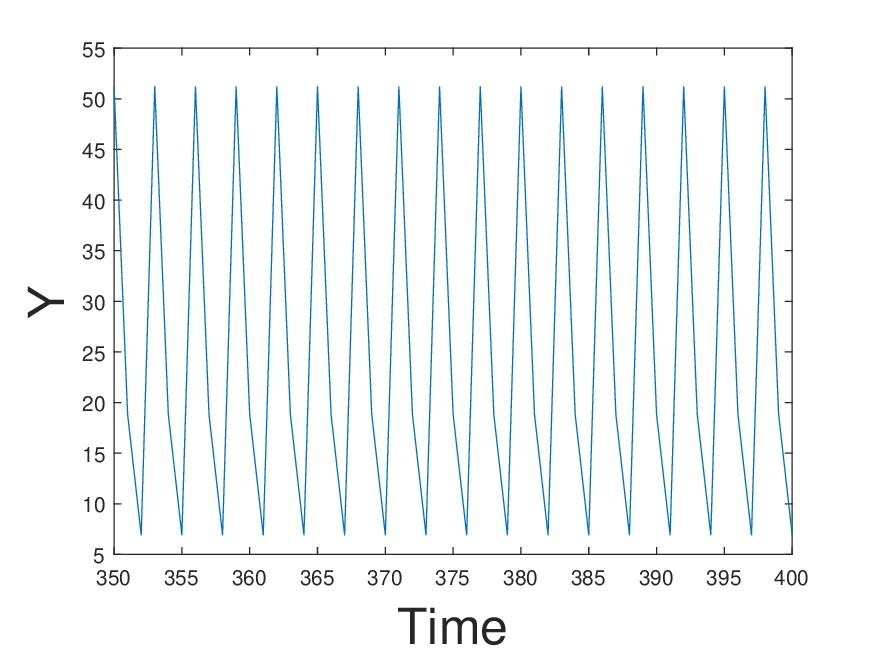}\\
		\vspace{0.02cm}
		\subcaption{}
	\end{minipage}%
	\begin{minipage}[t]{0.3\linewidth}
		\centering
		\includegraphics[width=1\linewidth]{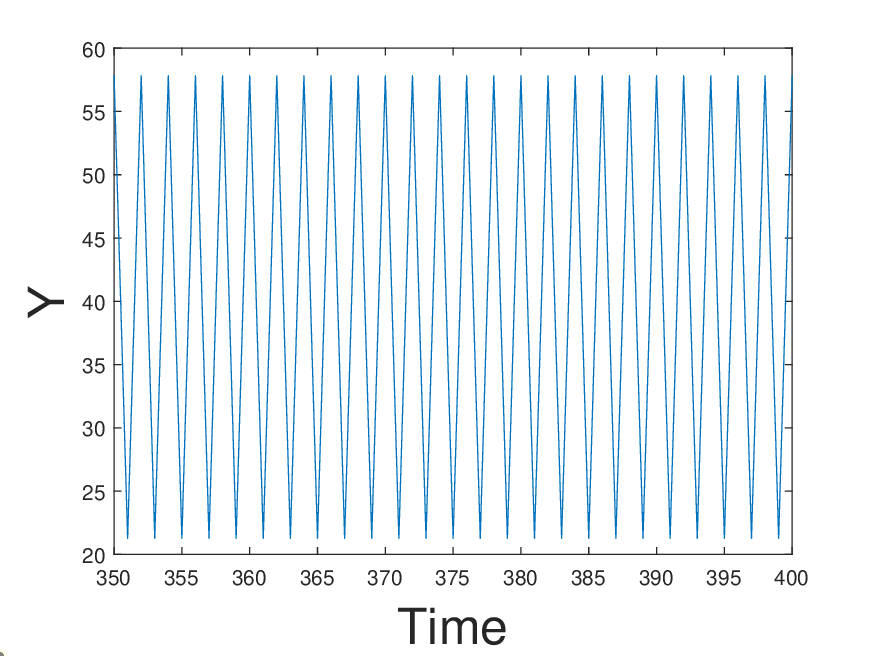}\\
		\vspace{0.02cm}
		\subcaption{}
		\centering
		\includegraphics[width=1\linewidth]{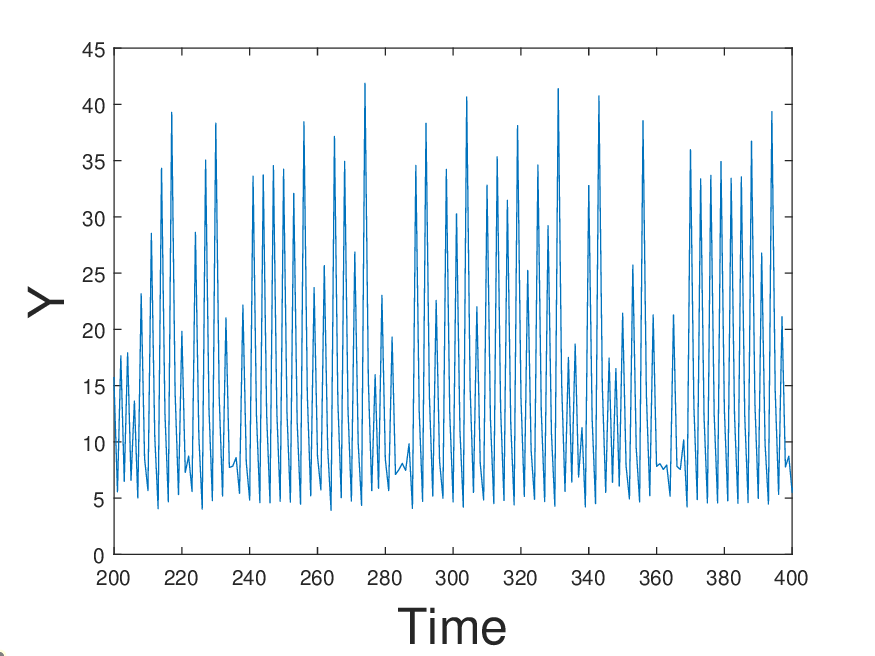}\\
		\vspace{0.02cm}
		\subcaption{}
	\end{minipage}%
	\centering
	\caption{CCNN neuron under different stimulus signals, the parameter is set to $ \alpha_{f}=0.1 $, $ \alpha_{e}=1 $, $ V_{E}=50 $. (a) $ S=0 $; (b) $ S=1 $; (c) $ S=3 $; (d) $ S=1+sin(t) $}
	\label{fig:CCNNsti}
\end{figure}

When the CCNN is driven by a periodic stimulus, it exhibits very complex dynamics, as shown in Fig. \ref{fig:CCNNsti}(d).
\par
\begin{figure}[!htb]
	\centering
	\includegraphics[width=0.4\linewidth]{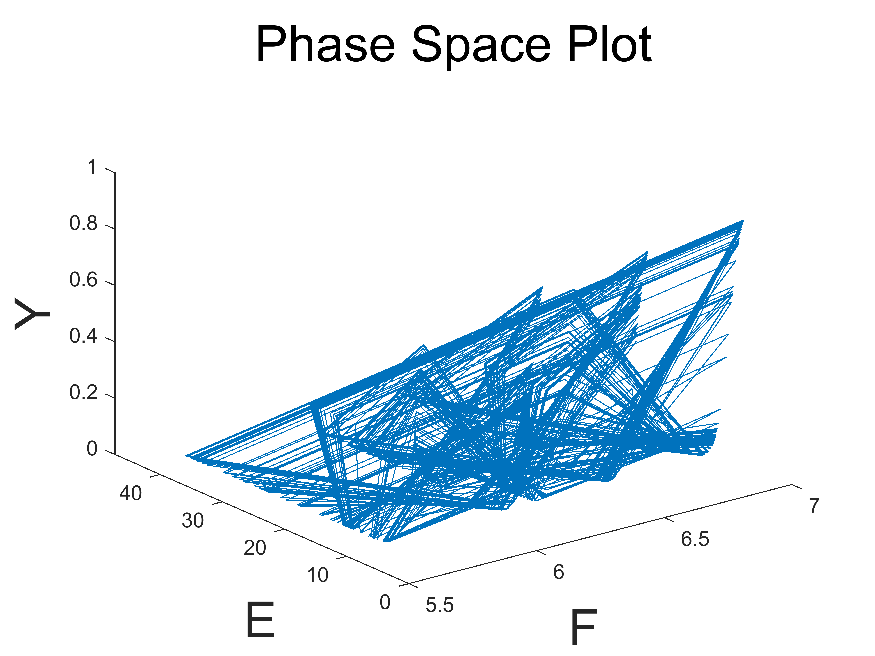}
	\caption{Phase-space plots of a single CCNN neuron, $ S=sin(t) $}
	\label{fig:PSP}
\end{figure}

A phase space is a space in which all possible states of a system are represented, with each possible state corresponding to one unique point in the phase space. We use values of $ F(t) $ as x-coordinate, values of $ E(t) $ as y-coordinate and values of $ Y(t) $ as z-coordinate to form phase space plots. The phase space plots of Fig. \ref{fig:CCNNsti}(d) are shown in Fig. \ref{fig:PSP}. The orbit of the CCNN does not repeat, diverge or converge, which indicates that the CCNN model exhibits chaotic behavior.
\par
\par
To further study this behavior, we calculated the Lyapunov exponent spectrum\cite{liu2018approach,liu2019novel}. When the largest Lyapunov exponent (LLE) is greater than zero, the system exhibits chaotic behavior. When the LLE is equal to zero, the system exhibits periodic behavior. When the LLE is less than zero, the trajectory of the system will converge to a stable fixed point.
\par
\begin{figure}[htbp]
	\centering
	\begin{minipage}[t]{0.3\linewidth}
		\centering
		\includegraphics[width=1\linewidth]{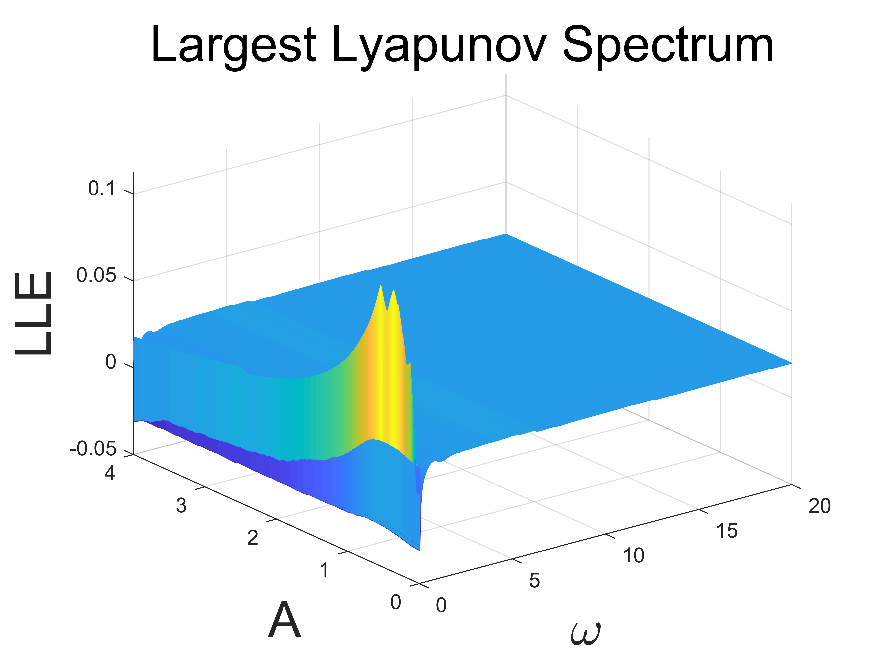}\\
		\vspace{0.02cm}
		\subcaption{$ S=A(1+sin(\omega t)) $}
	\end{minipage}
	\begin{minipage}[t]{0.3\linewidth}
		\centering
		\includegraphics[width=1\linewidth]{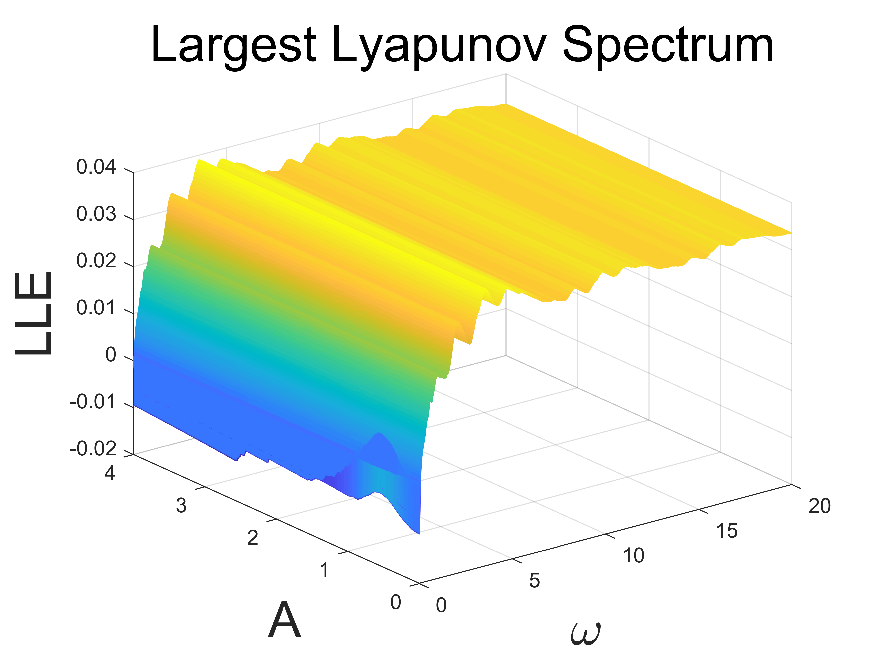}\\
		\vspace{0.02cm}
		\subcaption{$ S=Asin(\omega t) $}
	\end{minipage}%
	\centering
	\caption{Lyapunov exponent spectrum of single neuron under different stimulus signals.}
	\label{fig:LEs}
\end{figure}

 
When $ \alpha_{f}=0.1 $, $ \alpha_{e}=1.0 $, and $ V_{E}=50 $, $ S=0.5(1+sin(t)) $, the LLE converges to $ 0.09 $, which indicates that the neuron is in a chaotic state.
Furthermore, we used the Lyapunov exponent spectrum to investigate the overall dynamic behavior of single neuron under different stimulus signals. For $ S=A(1+sin(\omega t)) $, $ A\in [0,4] $, $ \omega\in [0,20] $, we calculate the corresponding LLE of CCNN.
The results are shown in Fig.\ref{fig:LEs}.
\par 
Clearly, the CCNN model can exhibit both periodic and chaotic behavior when stimulated by different driven signals. As shown in Fig. \ref{fig:LEs}, when $ S=A(1+sin(\omega t)) $, the CCNN model is sensitive to the amplitude of the driven signal. However, when $ S=A(sin(\omega t)) $, the CCNN is sensitive to the frequency of the driven signal. This interesting feature further empowers CCNN to generate a variety of response signals that reflect different input signals.
\subsection{Dynamic behavior of CCNN: Coupling with the neighborhood neurons}
When a CCNN neuron is coupled with neighborhood neurons, its equation is given by:
\begin{equation} \label{Eq:MultipleCCNN}
	\begin{split}
		\begin{aligned}
			F[n]&=e^{-\alpha_{f}}F[n-1]+S+\Sigma M\\
			L[n]&=e^{-\alpha_{l}}L[n-1]+\Sigma W\\
			U[n]&=F[n](1+\beta L[n])\\
			Y[n]&=1/(1+e^{-(U[n]-E[n])})\\
			E[n]&=e^{-\alpha_{e}}E[n-1]+V_{E}Y[n-1]
		\end{aligned}
	\end{split}
\end{equation}
where $ \Sigma M $ is the input feed from neighborhood neurons in the receptive field and $ \Sigma W $ is the coupling linking input of connected neurons.
\par
When $ S=C $, $ F[0]=0 $, $ L[0]=0 $,
\begin{equation} \label{Eq:MultipleCCNN1}
	\begin{split}
		\begin{aligned}
			F[n]&=e^{-\alpha_{f}}F[n-1]+C+ \Sigma M \\
			&=(C+\Sigma M)(1+e^{-\alpha_{e}}+e^{-2\alpha_{e}}+\cdots+e^{-(n-1)\alpha_{e}})\\
			&=(C+\Sigma M)(1-e^{-n\alpha_{e}})/(1-e^{\alpha_{e}})=C_{Sm}.
		\end{aligned}
	\end{split}
\end{equation}
\begin{equation} \label{Eq:MultipleCCNN2}
	\begin{split}
		\begin{aligned}
			L[n]&=e^{-\alpha_{f}}F[n-1]+\Sigma W \\
			&=\Sigma W(1+e^{-\alpha_{e}}+e^{-2\alpha_{e}}+\cdots+e^{-(n-1)\alpha_{e}})\\
			&=\Sigma W(1-e^{-n\alpha_{e}})/(1-e^{\alpha_{e}})=C_{Sw}
		\end{aligned}
	\end{split}.
\end{equation}
Therefore, Eq. (\ref{Eq:MultipleCCNN}) can be simplified to: 
\begin{equation} \label{Eq:MultipleCCNN3}
	\begin{split}
		\begin{aligned}
			Y[n]&=1/(1+e^{E[n]-C_{Sm}(1+\beta C_{Sw})})\\
			E[n]&=e^{-\alpha_{e}}E[n-1]+V_{E}Y[n-1]
		\end{aligned}
	\end{split}.
\end{equation}
Substituting $ Y[n-1] $ into $ E[n] $, we obtained the recursive formula of $ E_{n} $, which is:
\begin{equation} \label{Eq:MultipleCCNN4}
	\begin{split}
		\begin{aligned}
			E_{n}= e^{-\alpha_{e}}E[n-1]+V_{E}/(1+e^{E[n-1]-C_{Sm}(1+\beta C_{Sw})})
		\end{aligned}
	\end{split}.
\end{equation}
Therefore, the fixed points are:
\begin{equation} \label{Eq:MultipleCCNN5}
	\begin{split}
		\begin{aligned}
			E(1+e^{E-C_{Sm}(1+\beta C_{Sw})})=V_{E}/(1-e^{-\alpha_{e}})
		\end{aligned}
	\end{split}.
\end{equation}
The derivative function of the fixed point is given by: 
\begin{equation} \label{Eq:MultipleCCNN6}
	\begin{split}
		\begin{aligned}
			\frac{\partial f(E)}{\partial E}=e^{-\alpha_{e}}-V_{E}*\frac{e^{E-C_{Sm}(1+\beta C_{Sw})}}{(1+e^{E-C_{Sm}(1+\beta C_{Sw}})^{2}}
		\end{aligned}
	\end{split}.
\end{equation}
When $ E=ln0.5+C_{Sm}(1+\beta C_{Sw}) $, $ \frac{e^{E-C_{Sm}(1+\beta C_{Sw})}}{(1+e^{E-C_{Sm}(1+\beta C_{Sw})})^{2}} $ reaches its extreme point of $ 0.22 $. If $ \frac{1\pm e^{-\alpha_{e}}}{V_{E}}\leq 0.22 $, $ ln(-0.22V_{E}-1)\leq\mp\alpha_{e}\leq ln(0.22V_{E}-1) $, the fixed point is stable; otherwise, the fixed point is unstable.
\par
Let $ E_{n}=h(n) $; then, the derivative function of Eq. (\ref{Eq:singleCCNN6}) can be given as follows:
\begin{equation} \label{Eq:MultipleCCNN7}
	\begin{split}
		\begin{aligned}
			\frac{\partial f(E)}{\partial n}&=\frac{\partial f(E)}{\partial E}*\frac{\partial E}{\partial n}=\frac{\partial f(E)}{\partial E}*\frac{\partial h(n)}{\partial n}\\
			&=(e^{-\alpha_{e}}-V_{E}*\frac{e^{E-C_{Sm}(1+\beta C_{Sw})}}{(1+e^{E-C_{Sm}(1+\beta C_{Sw}})^{2}})*\frac{\partial h(n)}{\partial n}
		\end{aligned}
	\end{split}.
\end{equation}
\par
If $ f^{T}(E)=E $, then $ T $ is the period of $ E $. Since $ e^{E-C_{Sm}(1+\beta C_{Sw})}>0 $, $ g(C_{S})=e^{-\alpha_{e}}-V_{E}*\frac{e^{E-C_{Sm}(1+\beta C_{Sw})}}{(1+e^{E-C_{Sm}(1+\beta C_{Sw})})^{2}} $ is a monotonically increasing function. Since $ Y[n]=1/(1+e^{-(U[n]-E[n])})>0 $, $ \Sigma M $ and $ \Sigma W $ are always positive. Therefore, when a neuron is coupled with its neighborhood neurons, it will have a higher output signal frequency.
\section{Spiking Feature of CCNN}
\begin{figure}[!htb]
	\centering
	\includegraphics[width=0.7\linewidth]{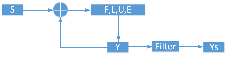}
	\caption{Flowchart for the CCNN sequence of generating spike train}
	\label{fig:threshold filter}
\end{figure}

To evaluate the spiking feature of the CCNN, we utilized a threshold filter. The flowchart is shown in Fig. \ref{fig:threshold filter}.
\par
\begin{figure}[htbp]
	\centering
	\begin{minipage}[t]{0.3\linewidth}
		\centering
		\includegraphics[width=1\linewidth]{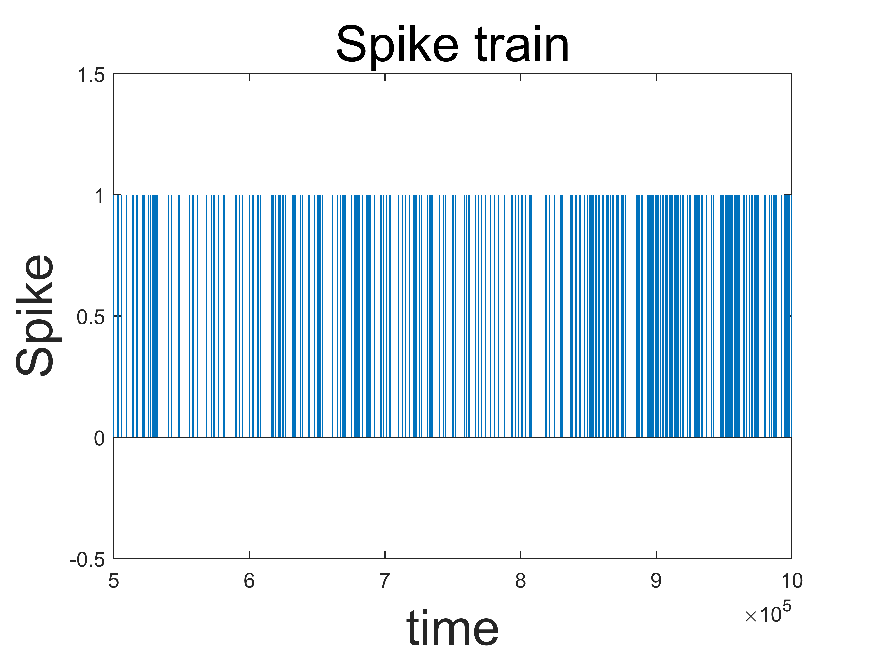}\\
		\subcaption{}
		\vspace{0.02cm}
	\end{minipage}
	\begin{minipage}[t]{0.3\linewidth}
		\centering
		\includegraphics[width=1\linewidth]{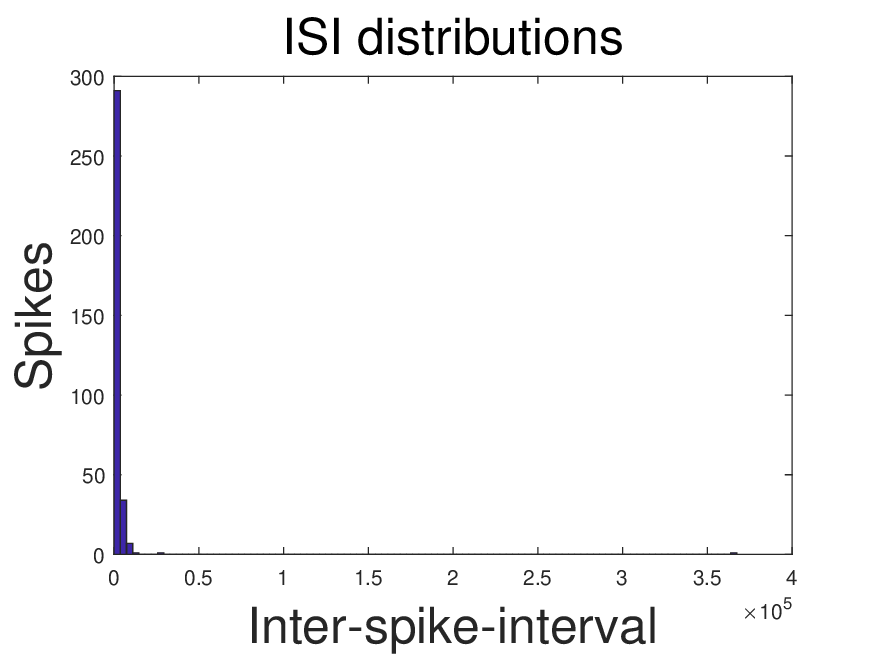}\\
		\subcaption{}
		\vspace{0.02cm}
	\end{minipage}%
	\centering
	\caption{Spike train and ISI distributions under a stimulus of $ y=0.21*(square(\omega*t,dc)+1) $, where $ \omega=0.2\pi $ and $ dc=50 $}
	\label{fig:CCNN_monkey}
\end{figure}

In Fig. \ref{fig:threshold filter} the threshold filter is designed to be: 
\begin{equation} \label{Eq:filter}
	\begin{split}
		\begin{aligned}
			Y_{S}[n]&=\left\{
			\begin{aligned}
				1, & \qquad if \qquad Y_{S}[n]>\mu max(Y)\\
				0, & \qquad otherwise \\
			\end{aligned}
			\right. \\
		\end{aligned}
	\end{split},
\end{equation}
\par
where $ \mu $ is a coefficient. When $ \mu =0.8 $, $ S=0.21*(square(\omega*t,dc)+1) $ with angular velocity $ \omega=0.2\pi $ and duty cycle $ dc=50 $. The spike train and ISI distributions are shown in Fig. \ref{fig:CCNN_monkey}.
\par
Comparing Fig. \ref{fig:CCNN_monkey}(b) with Fig. \ref{fig:RealISI} indicates that the CCNN model can generate a pattern similar to that of real neurons.
\par
Siegel performed detailed experiments of real signal neurons of the primary visual cortex and reported that "Stimuli were presented at 271.5 ms intervals. A burst of action potentials can be seen following each stimulus. By plotting the distribution of interspike intervals, it can be seen that there is a major periodicity at approximately 270 ms. There are also peaks at 175, 540, and 1080 ms. The latter two values are roughly integer multiples of the driving period" \cite{siegel1990non}.
\par
\begin{figure}[htbp]
	\centering
	\begin{minipage}[t]{0.3\linewidth}
		\centering
		\includegraphics[width=1\linewidth]{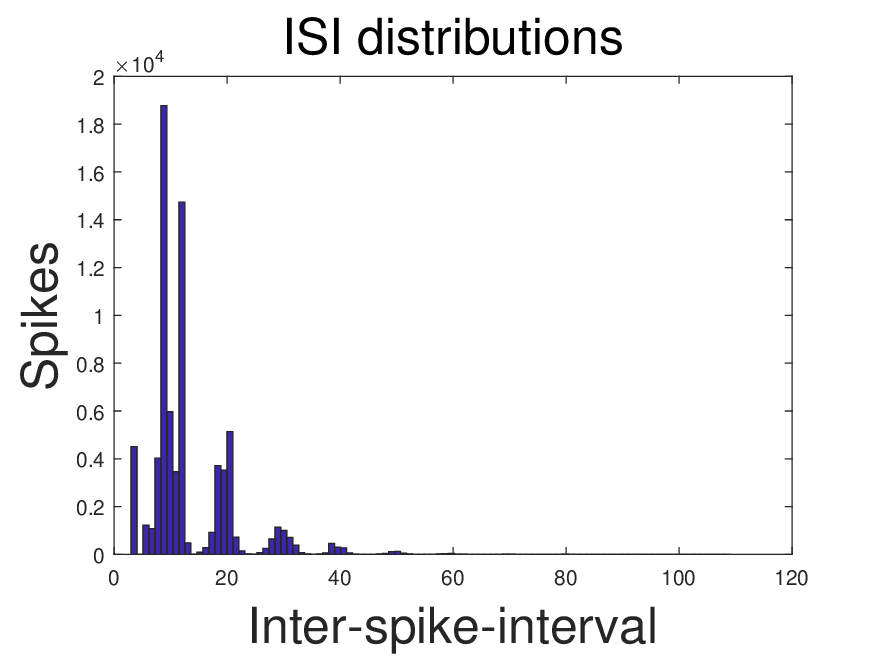}\\
		\vspace{0.02cm}
	\end{minipage}
	\begin{minipage}[t]{0.3\linewidth}
		\centering
		\includegraphics[width=1\linewidth]{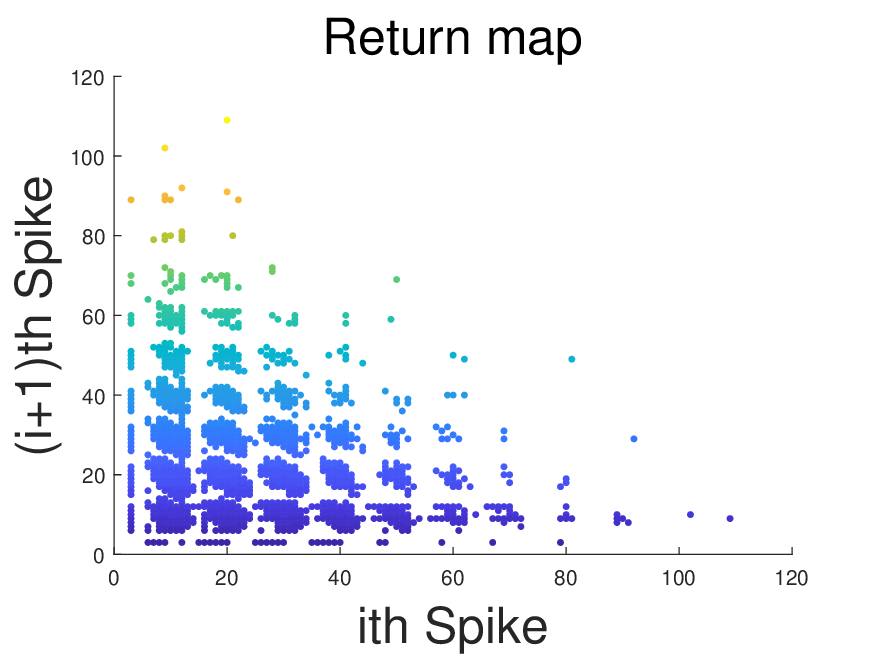}\\
		\vspace{0.02cm}
	\end{minipage}%
	\centering
	\caption{ ISI distributions and return map of the i-th spike - (i+1)-th spike under the stimulus of $ y=0.4*(square(\omega*t,dc)+1) $, where $ \omega=0.2\pi $ (driven period is $ T=10 $) and $ dc=50 $}
	\label{fig:ISIandReturnmap of siegel}
\end{figure}

The CCNN model can reproduce this phenomenon perfectly. When the stimulus is $ y=0.4*(square(\omega*t,dc)+1) $, where $ \omega=0.2\pi $ (driven period is $ T=10 $) and $ dc=50 $, the ISI distributions and return are shown in Fig. \ref{fig:ISIandReturnmap of siegel}.
\par
Furthermore, Siegel showed that a single neuron of the primary visual cortex can exhibit different patterns under different stimulus frequencies (see Fig. 3~Fig. 5 in ref\cite{siegel1990non}). We used different frequencies of the driven signal to stimulate the CCNN model and also found that different patterns of spike trains exist.
\par
\begin{figure*}[htbp]
	\centering
	\begin{minipage}[t]{0.3\linewidth}
		\centering
		\includegraphics[width=1\linewidth]{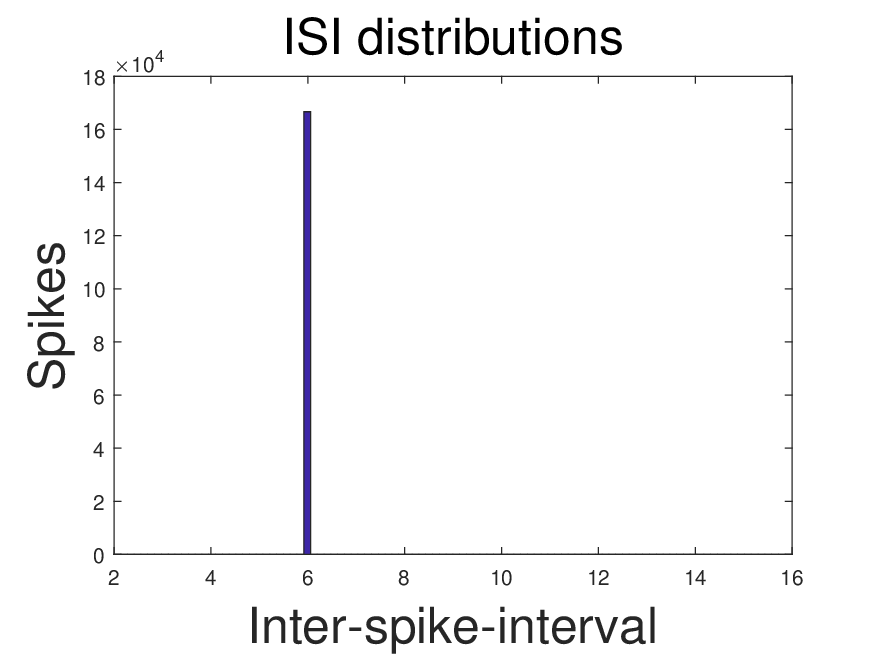}\\
		\vspace{0.02cm}
		\subcaption{}
		\includegraphics[width=1\linewidth]{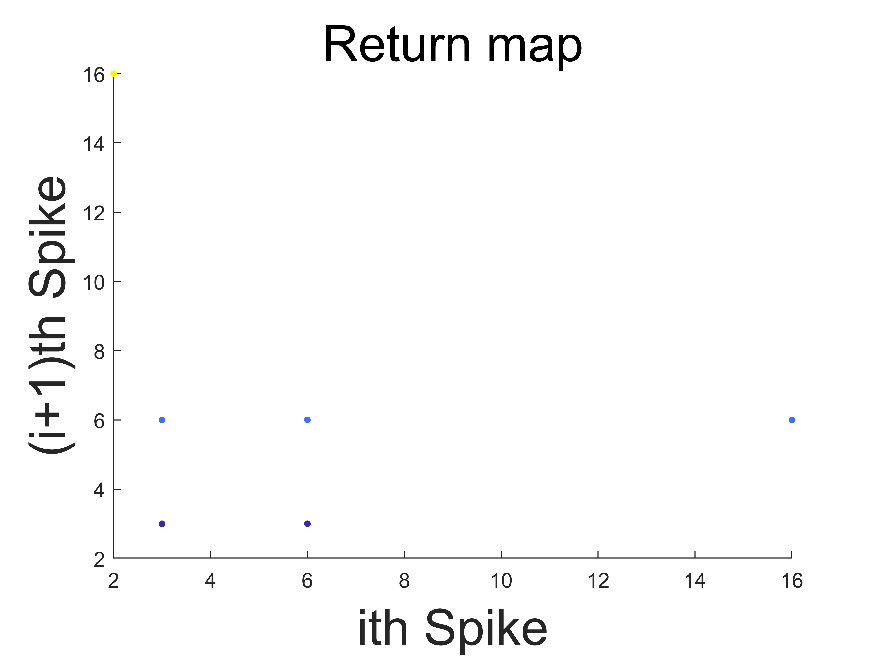}\\
		\vspace{0.02cm}
		\subcaption{}
	\end{minipage}%
	\begin{minipage}[t]{0.3\linewidth}
		\centering
		\includegraphics[width=1\linewidth]{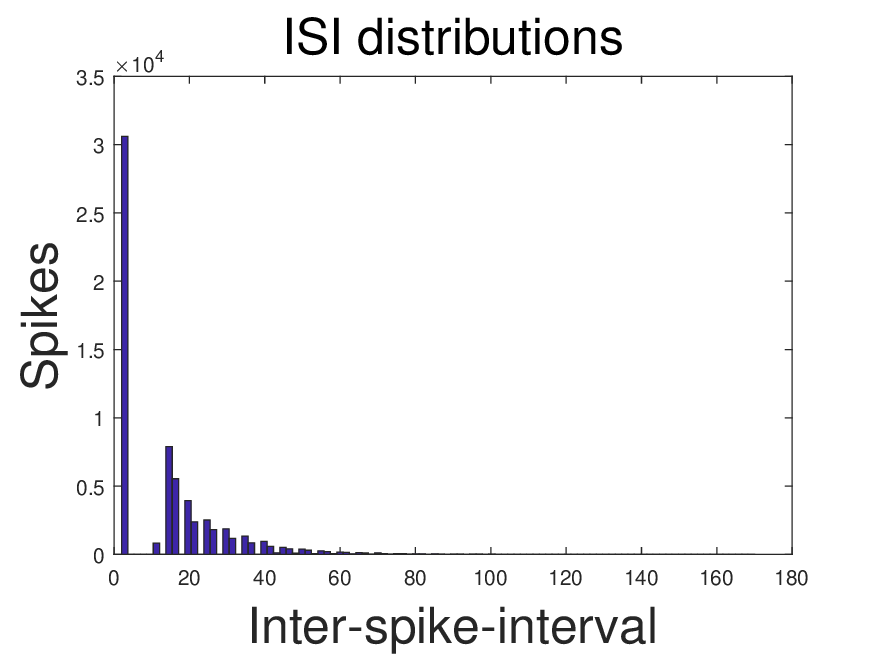}\\
		\vspace{0.02cm}
		\subcaption{}
		\includegraphics[width=1\linewidth]{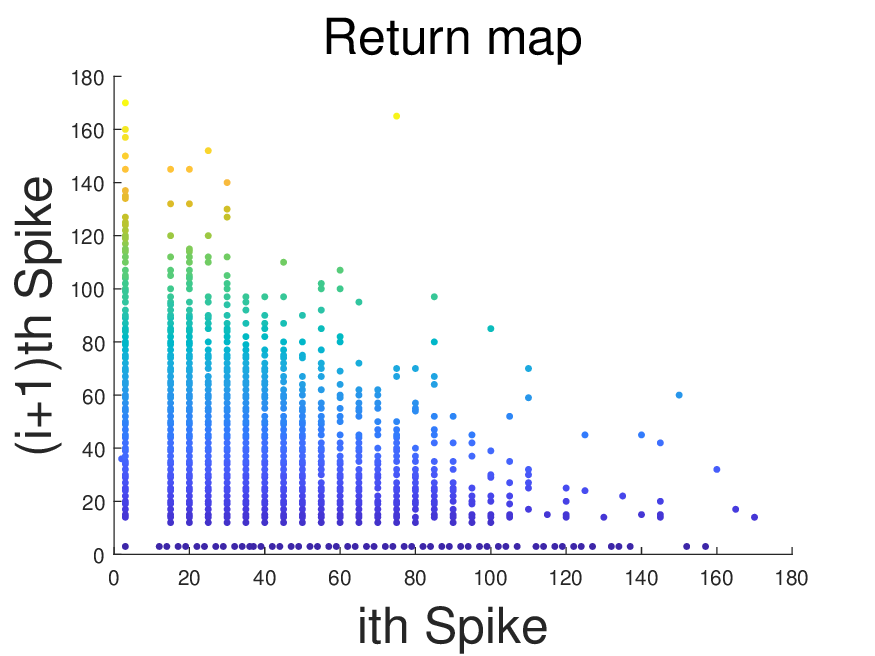}\\
		\vspace{0.02cm}
		\subcaption{}
	\end{minipage}%
	\begin{minipage}[t]{0.3\linewidth}
		\centering
		\includegraphics[width=1\linewidth]{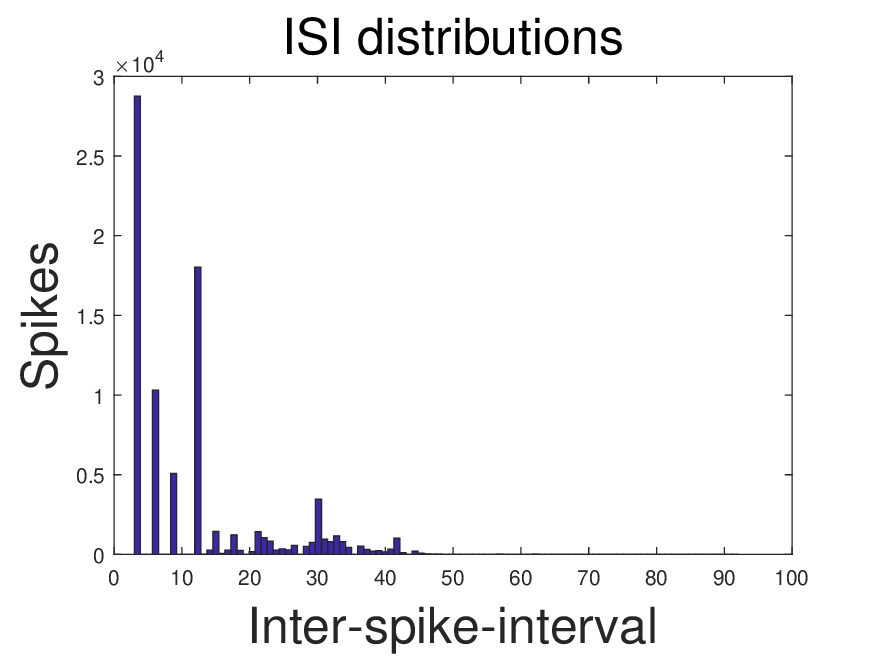}\\
		\vspace{0.02cm}
		\subcaption{}
		\includegraphics[width=1\linewidth]{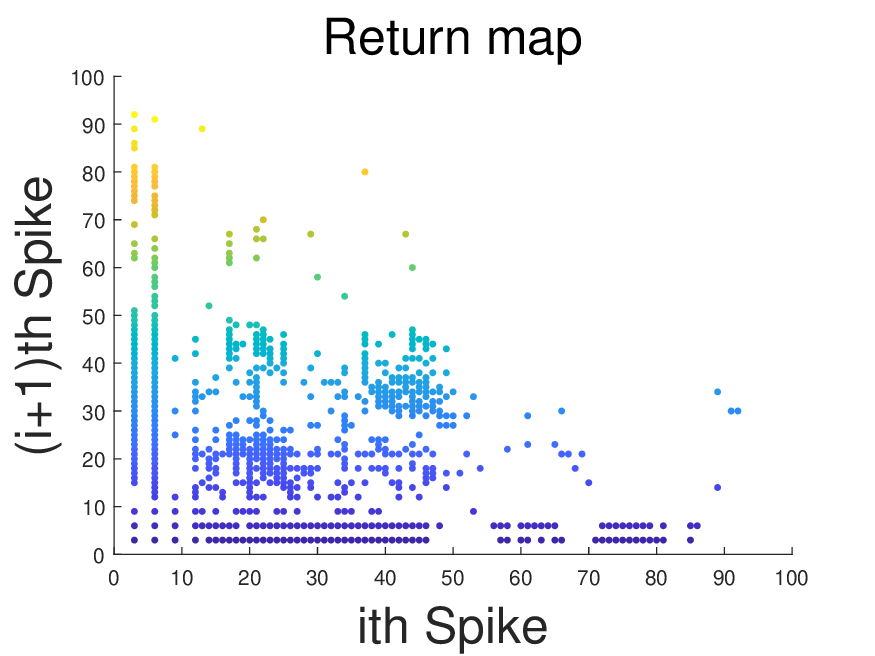}\\
		\vspace{0.02cm}
		\subcaption{}
	\end{minipage}%
	\centering
	\caption{Spike train of CCNN under periodic stimulus signal: (a) shows the ISI distributions under the stimulus of $ y=0.21*(square(\omega*t,dc)+1) $, where $ \omega=2\pi $ and $ dc=50 $; (b) is the return map under the stimulus of $ y=0.21*(square(\omega*t,dc)+1) $, where $ \omega=2\pi $ and $ dc=50 $; (c) shows the ISI distributions under the stimulus of $ y=0.21*(square(\omega*t,dc)+1) $, where $ \omega=1.2\pi $ and $ dc=50 $; (d) is the return map under the stimulus of $ y=0.21*(square(\omega*t,dc)+1) $, where $ \omega=1.2\pi $ and $ dc=50 $; (e) shows the ISI distributions under the stimulus of $ y=0.21*(square(\omega*t,dc)+1) $, where $ \omega=\pi $ and $ dc=50 $; (f) is the return map under the stimulus of $ y=0.21*(square(\omega*t,dc)+1) $, where $ \omega=\pi $ and $ dc=50 $;}
	\label{fig:different stimulus of CCNN}
\end{figure*}

When the stimulus signal is set to $ y=0.21*(square(\omega*t,dc)+1) $, $ \omega=2\pi $ is the angular velocity and $ dc=50 $ is the duty cycle. The spike train and ISI distributions are shown in Fig.\ref{fig:different stimulus of CCNN} (a) and (b), indicating that neurons exhibit periodic behavior.
\par
When the frequency of the stimulus signal is decreased to $ \omega=1.2\pi $, the neuron starts to exhibit complex behavior. The main frequency is equal to the stimulus signal's frequency, but frequencies of four to forty-four times the stimulus signal frequency can be observed, and their intensity decreases with increasing frequency. Small windows with frequencies of 1 and 4 times the frequency of the stimulus signal exist; however, the neuron does not output double and triple frequency signals, as shown in Fig.\ref{fig:different stimulus of CCNN} (c) and (d).
\par
When the frequency of the stimulus signal is decreased to $ \omega=1\pi $, the neuron exhibits a complex temporal pattern, as shown in Fig.\ref{fig:different stimulus of CCNN} (e) and (f).
\par
\par
The results of the above experiments demonstrate that the CCNN model is more similar to a real primary visual cortex neuron network.
\par
The ultimate advantage of SNNs comes from their ability to fully exploit spatiotemporal event-based information. Using CCNN as a unit neuron allows richer spike trains to be output from different input signals. These spike trains can be used as a type of "error" that can be backpropagated by algorithms. Therefore, it is possible to utilize CCNN to design deep learning networks for artificial intelligence purposes.
\section{Applications in Image Segmentation}
\begin{figure}[!htb]
	\centering
	\includegraphics[width=0.4\linewidth]{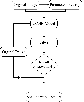}
	\caption{Flowchart of image segmentation}
	\label{fig:ImageSegmentationFlowChart}
\end{figure}

Since CCNN model is more similar to the visual cortex, it is natural to utilize this model to perform image processing tasks. In this work, we use applied CCNN model into a image segmentation scheme. The flowchart is shown in Fig. \ref{fig:ImageSegmentationFlowChart}.  
\par
Where the filter is designed as follows:
\begin{equation} \label{Eq:ipfilter1}
	\begin{split}
		\begin{aligned}
			Y_{S}[n]&=\left\{
			\begin{aligned}
				1, & \qquad if \qquad Y_{S}[n]>\mu max(I)\\
				0, & \qquad otherwise \\
			\end{aligned}
			\right. \\
		\end{aligned}
	\end{split},
\end{equation}
\par
where $ \mu $ is a coefficient set to $ 0.33 $ in this application, and $ max(I) $ is the largest pixel value of the image.
\par  
The main image segmentation procedures are described as follows:
\par
Step 1: Generate the neural network which has the same size of the input image.
\par
Step 2: Calculate standard deviation and otsu threshold of the input image and set the exponential decay factors based on them. Set the initial state of all neurons. The initial state is arbitrary, in this application we set all initial state to be zero for simplicity,
\par
Step 3: Set coupling link mode. In this application, we set the coupling link mode so that each neuron only connects to the nearest 8 neurons around it. The linking weight is set to be the reciprocal of the Euclidean distance. For neuron $ N(i,j) $ and  $ N(k,l) $, the linking weight is $ M_{ijkl},W_{ijkl}=\frac{1}{(i-k)^{2}+(j-l)^{2}} $. 
\par 
Step 4: Using input image as the feeding input $ F $, calculate each neuron's modulation product $ U $, dynamic threshold $ E $ and output $ Y $ and $ Y_{S} $
Step 5: Repeat step 4 until the results is converged.
Step 6: Using the final $ Y_{S} $ to generate the segmentation image. 
\par 
The pseudocode of the algorithm is shown in Algorithm \ref{alg:b}.
\begin{algorithm}[h]
	\caption{Algorithm for Image Segmentation}
	\label{alg:b}
	\begin{algorithmic}
		\REQUIRE Original image (OI)
		\ENSURE Segmentation result (SR)
		\STATE $ theta = std2(OI) $
		\STATE $ Shist = gh(OI) $
		\STATE $ Smax=max(max(OI)) $
		\STATE $ af=log(1/theta) $
		\STATE $ vl=1 $
		\STATE $ ve=exp(-af)+1+6*B*vl $
		\STATE $ M3=(1-exp(-(3*af)))/(1-exp(-af))+6*B*vl*exp(-af) $
		\STATE $ ae=log(ve/(Shist*M3)) $
		\STATE $ W=\begin{bmatrix}
			0.5 & 1 & 0.5 \\
			1 & 0 & 1 \\
			0.5 & 1 & 0.5
		\end{bmatrix} $
		\STATE $ W=M $
		\STATE $ Y=0 $
		\STATE $ L=0 $
		\STATE $ U=0 $
		\STATE $ E=0 $
		\REPEAT
		\STATE $ K=conv2(Y,M,'same') $
		\STATE $ F=OI $
		\STATE $ L=vl*K $
		\STATE $ U=exp(-af)*U+F*(1+B*L) $
		\STATE $ Y=1/(1+exp(E-U)) $
		\STATE $ E=exp(-ae)*E+ve*Y $
		\IF{$Y>\mu *max(OI)$)}
		\STATE $ Y=1 $
		\ELSE
		\STATE $ Y=0 $
		\ENDIF
		\STATE $SR \leftarrow Y$
		\UNTIL{ Segmentation result is converged }
		\RETURN Segmentation result (SR)
	\end{algorithmic}
\end{algorithm}
\par
\begin{figure}[htbp]
	\centering
	\begin{minipage}[t]{0.2\linewidth}
		\centering
		\includegraphics[width=0.75\linewidth,height=1\linewidth]{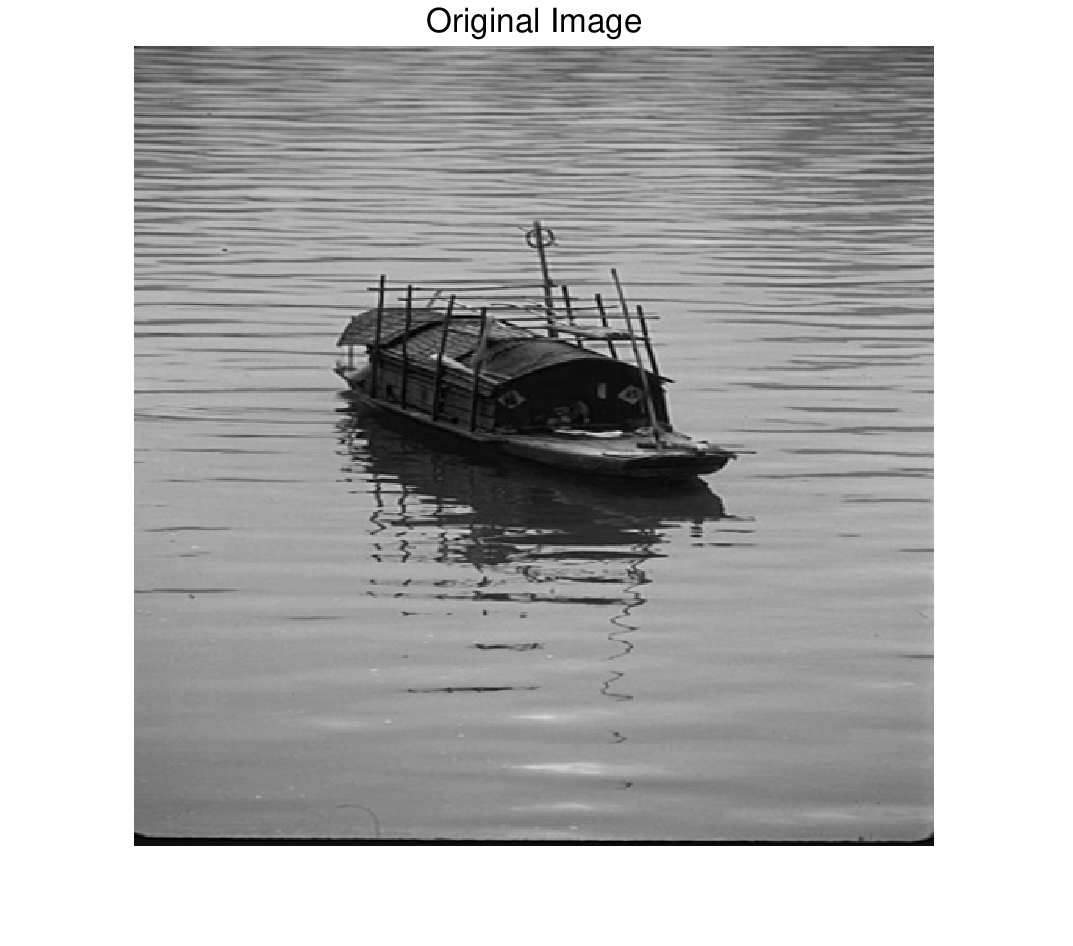}\\
		\vspace{0.02cm}
		\subcaption{}
		\includegraphics[width=0.75\linewidth,height=1\linewidth]{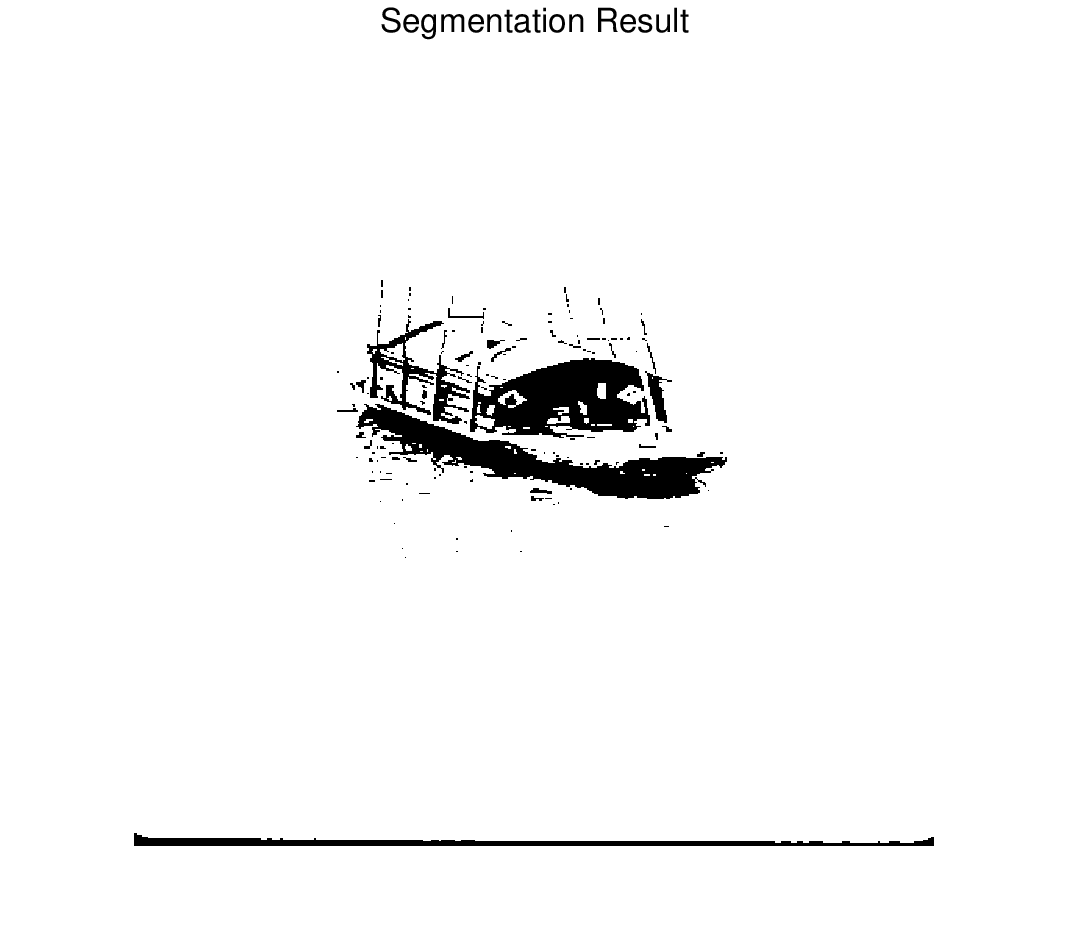}\\
		\vspace{0.02cm}
		\subcaption{}
	\end{minipage}%
	\begin{minipage}[t]{0.2\linewidth}
		\centering
		\includegraphics[width=0.75\linewidth,height=1\linewidth]{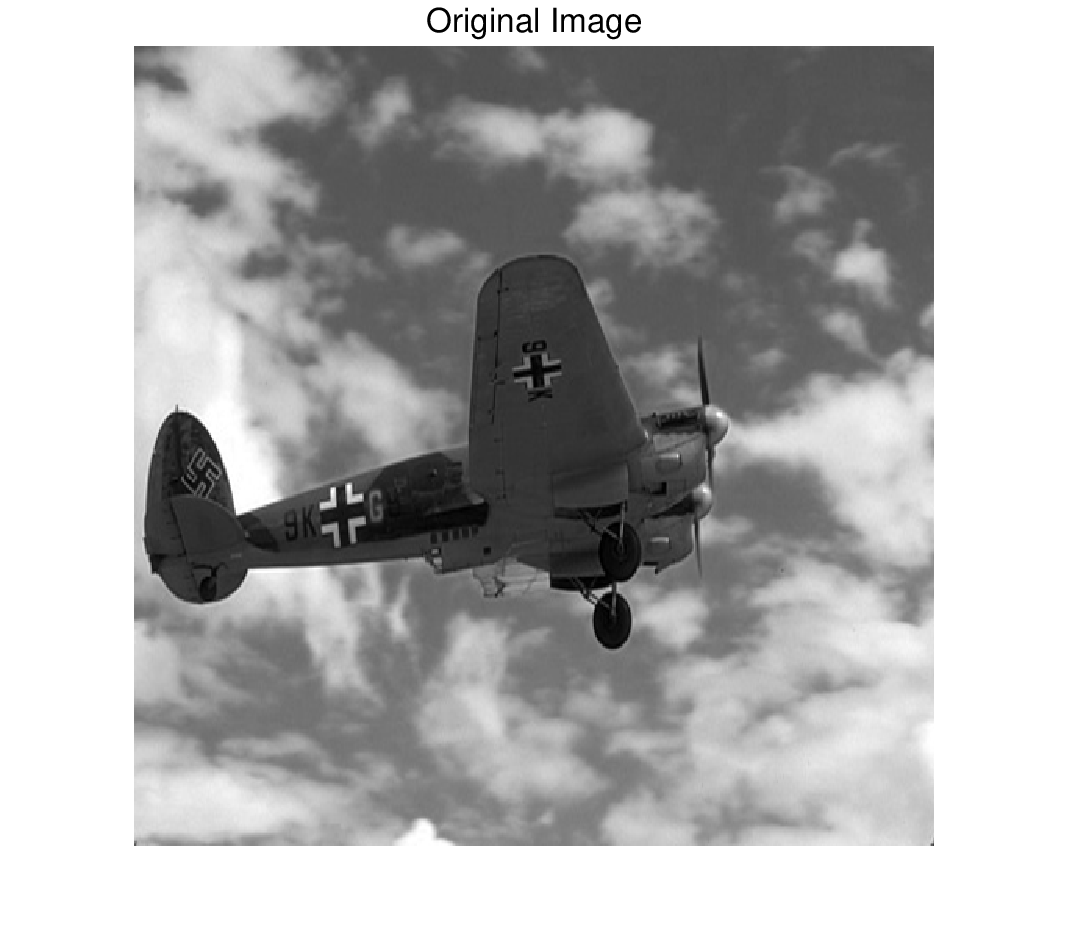}\\
		\vspace{0.02cm}
		\subcaption{}
		\includegraphics[width=0.75\linewidth,height=1\linewidth]{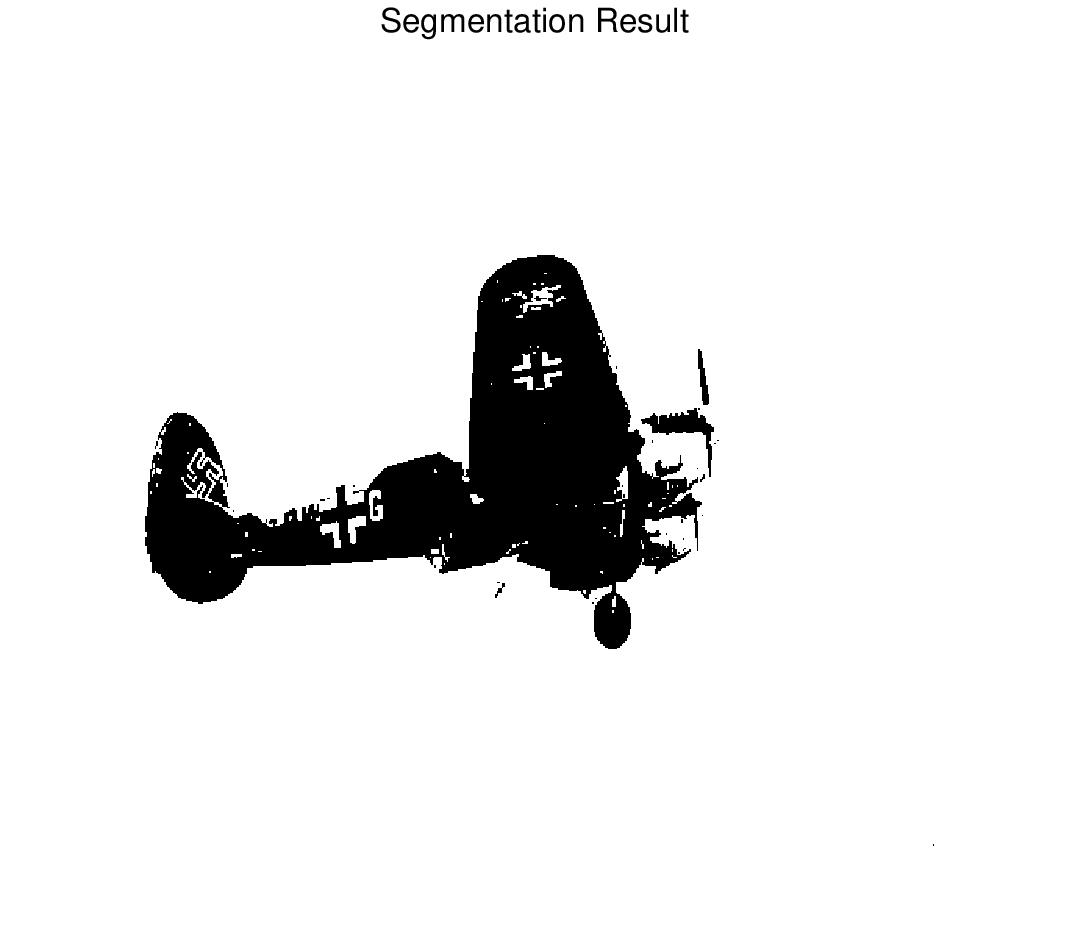}\\
		\vspace{0.02cm}
		\subcaption{}
	\end{minipage}%
	\begin{minipage}[t]{0.2\linewidth}
		\centering
		\includegraphics[width=0.75\linewidth,height=1\linewidth]{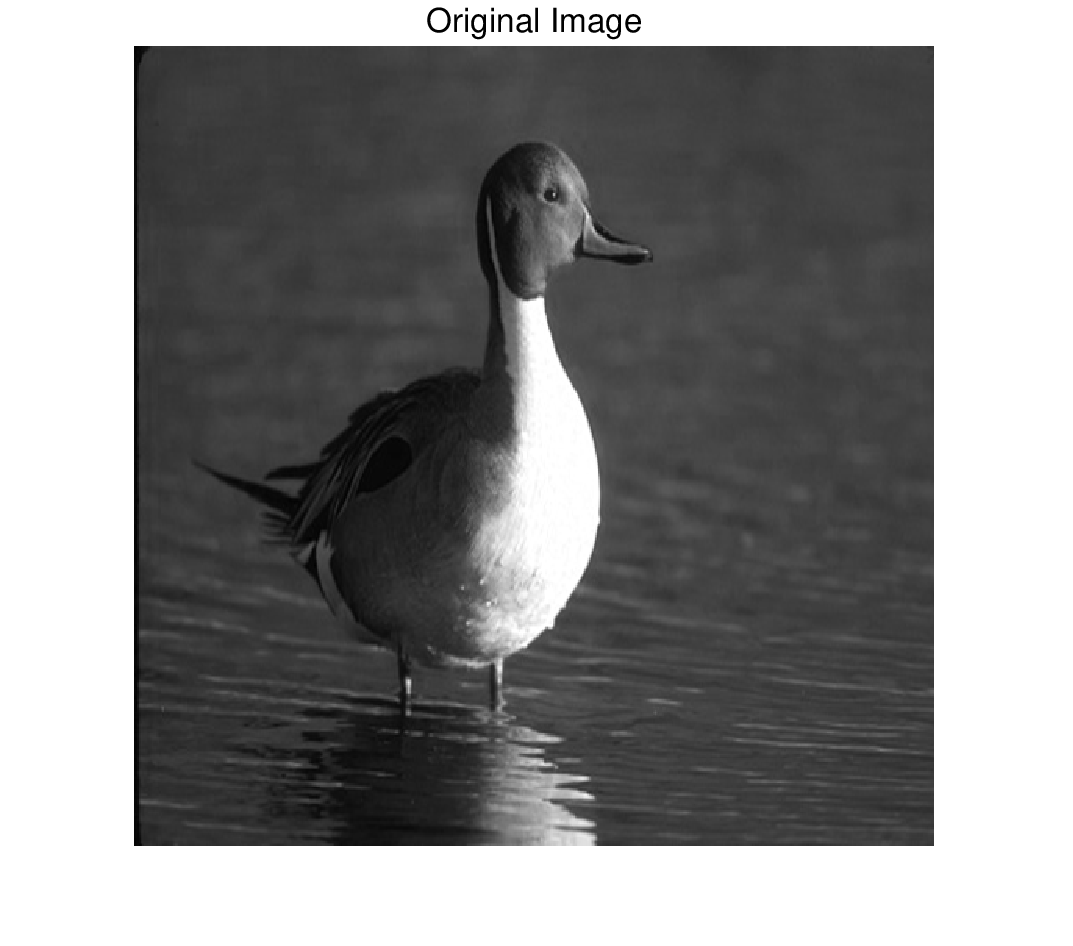}\\
		\vspace{0.02cm}
		\subcaption{}
		\includegraphics[width=0.75\linewidth,height=1\linewidth]{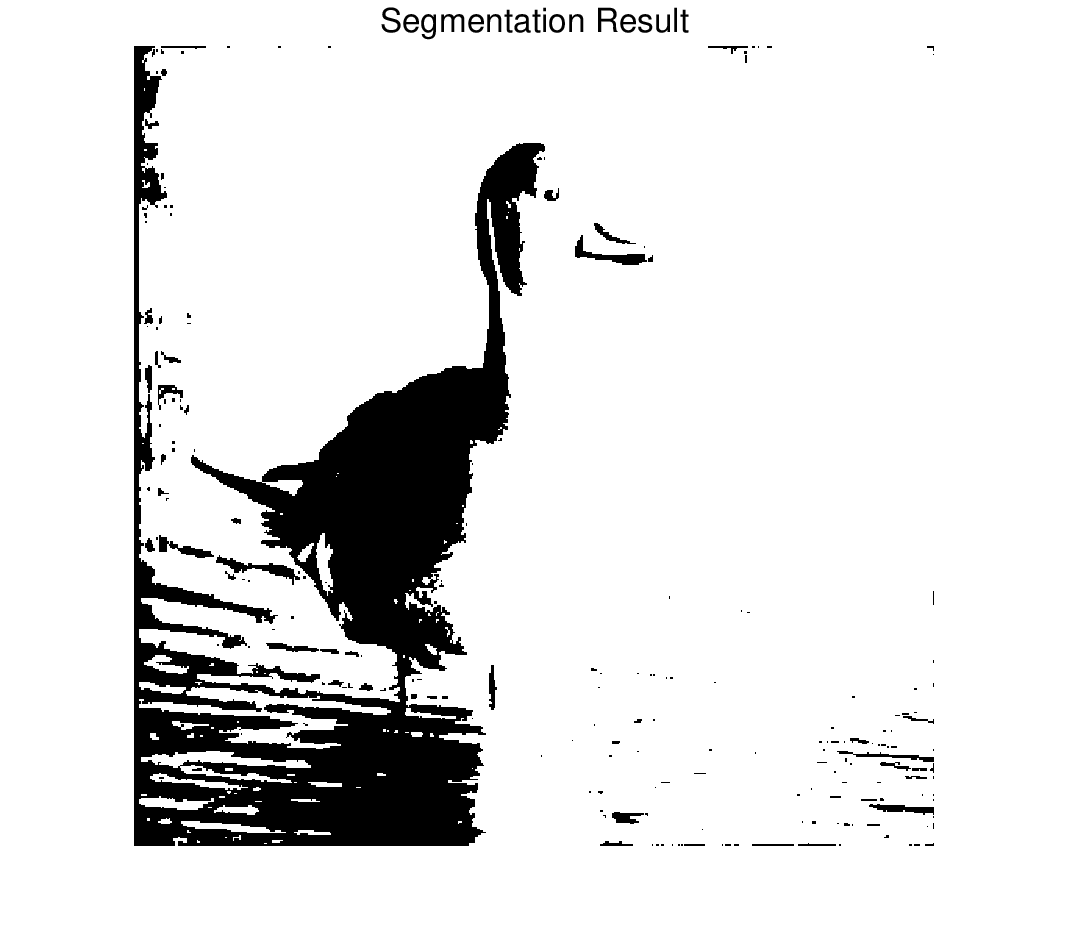}\\
		\vspace{0.02cm}
		\subcaption{}
	\end{minipage}
	\centering
	\caption{Natural image segmentation results: (a) original image of boat; (b) segmentation results of boat; (c) original image of plane; (d) segmentation results of plane; (e) original image of duck; (f) segmentation results of duck;}
	\label{fig:nature SR experiments}
\end{figure}

To evaluate the effectiveness of the proposed segmentation method, we use BSDS500 dataset, which is designed for evaluating natural edge detection that includes not only object contours but also background boundaries. The visual segmentation results are shown in Fig.\ref{fig:nature SR experiments}.
\par
The results of Fig.\ref{fig:nature SR experiments}(a)--Fig.\ref{fig:nature SR experiments}(d) have shown that the proposed method can effectively extract the target object from the background. However, our method is an unsupervised image segmentation method and depends on the difference of pixel values of the target object and the background. Therefore, our method cannot distinguish the object which has different pixel values, as shown in Fig.\ref{fig:nature SR experiments}(e) and Fig.\ref{fig:nature SR experiments}(f).
\par 
Different from natural images, medical images use the different characteristics of normal and abnormal tissues in X-rays (CT images) or magnetic fields (MRI images) to identify the lesion area. In medical images, if the pixel value of one area is obviously different from other areas, then the area is likely to be the lesion area. Therefore, our method can effectively perform medical image segmentation tasks. 
\par 
To evaluate the effectiveness of the proposed segmentation method, we conducted experiments on images from the Mammographic Image Analysis Society (MIAS) database. For this database, the filter parameters $ \mu $ is set to be $ 0.45 $.  We selected different types of breast images; the details are shown in Tab. \ref{tab:breast image}.
\begin{table}[htbp] 
	\centering
	{\tiny \begin{tabular}{lccl}
			\toprule
			Reference Number & Background Tissue & Class of abnormality present & Severity of abnormality \\
			\midrule
			mdb007 & Fatty-glandular & Normal & N/A\\
			mdb030 & Fatty-glandular & Other, ill-defined masses & Benign\\
			mdb158 & Fatty & Architectural distortion & Malignant\\
			mdb120 & Fatty-glandular & Architectural distortion & Malignant\\
			mdb121 & Fatty-glandular & Architectural distortion & Benign\\
			mdb238 & Fatty & Calcification & Malignant\\
			\bottomrule
	\end{tabular}}
	\caption{The breast image types}
	\label{tab:breast image}
\end{table}
\par
\begin{figure*}[htbp]
	\centering
	\begin{minipage}[t]{0.15\linewidth}
		\centering
		\includegraphics[width=0.75\linewidth,height=1\linewidth]{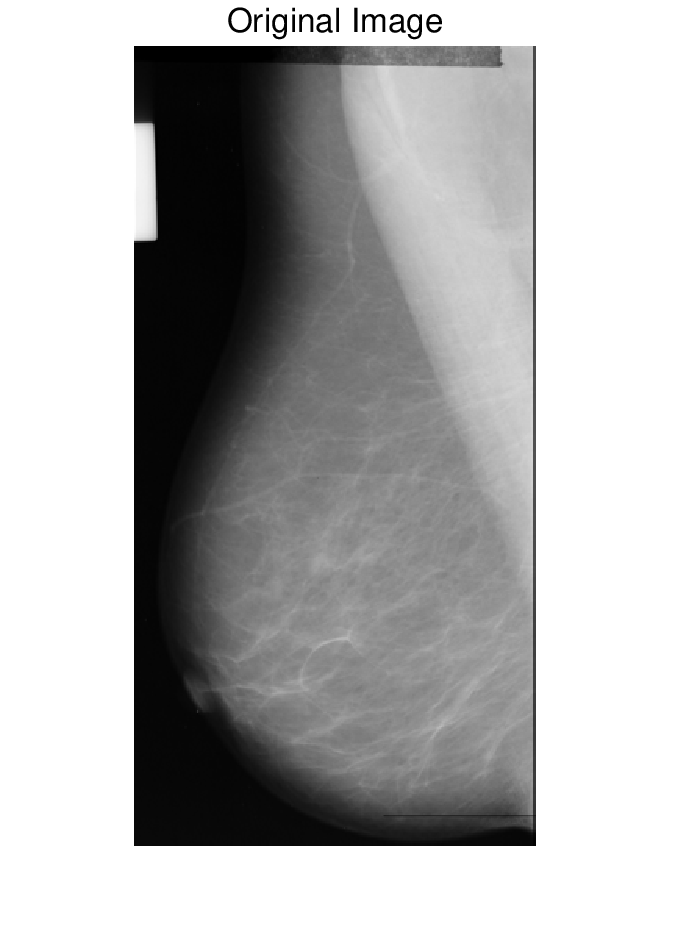}\\
		\vspace{0.02cm}
		\subcaption{}
		\includegraphics[width=0.75\linewidth,height=1\linewidth]{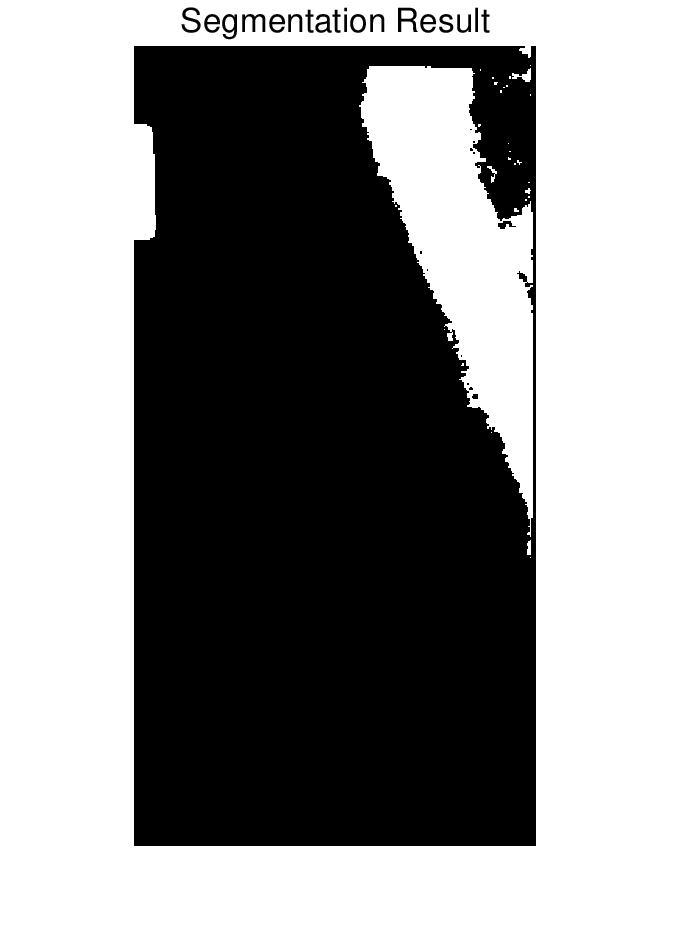}\\
		\vspace{0.02cm}
		\subcaption{}
	\end{minipage}%
	\begin{minipage}[t]{0.15\linewidth}
		\centering
		\includegraphics[width=0.75\linewidth,height=1\linewidth]{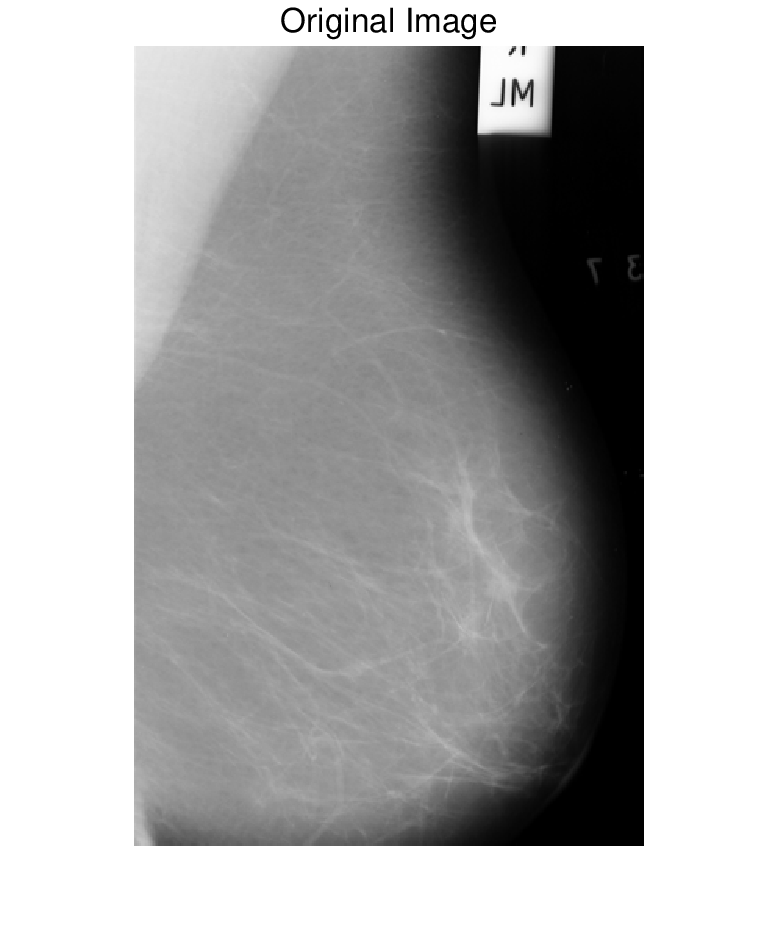}\\
		\vspace{0.02cm}
		\subcaption{}
		\includegraphics[width=0.75\linewidth,height=1\linewidth]{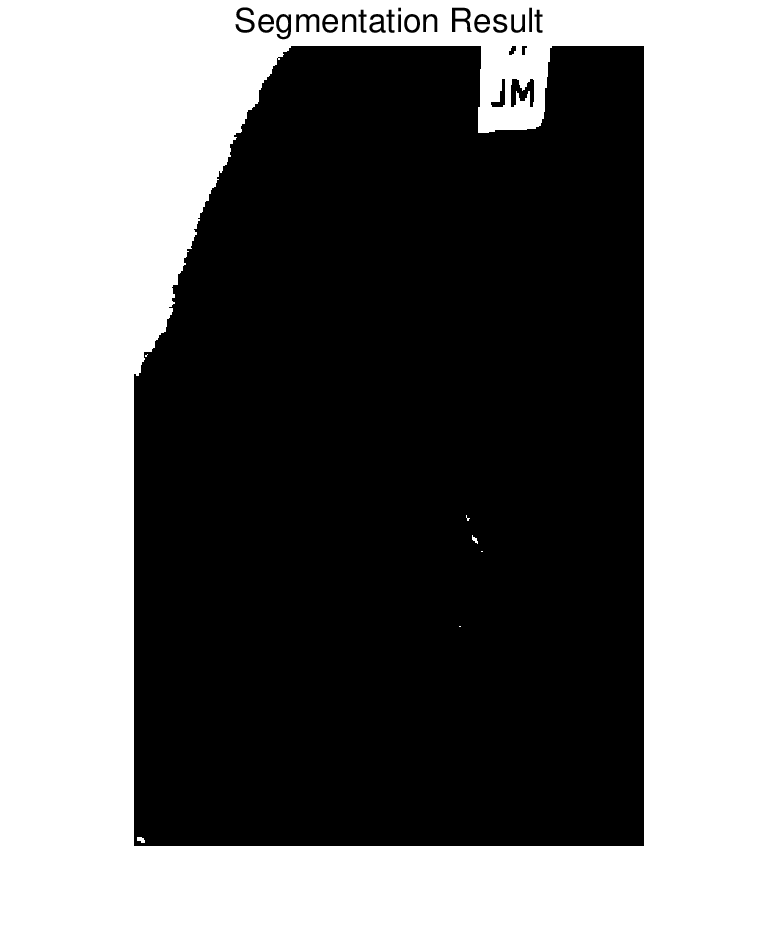}\\
		\vspace{0.02cm}
		\subcaption{}
	\end{minipage}%
	\begin{minipage}[t]{0.15\linewidth}
		\centering
		\includegraphics[width=0.75\linewidth,height=1\linewidth]{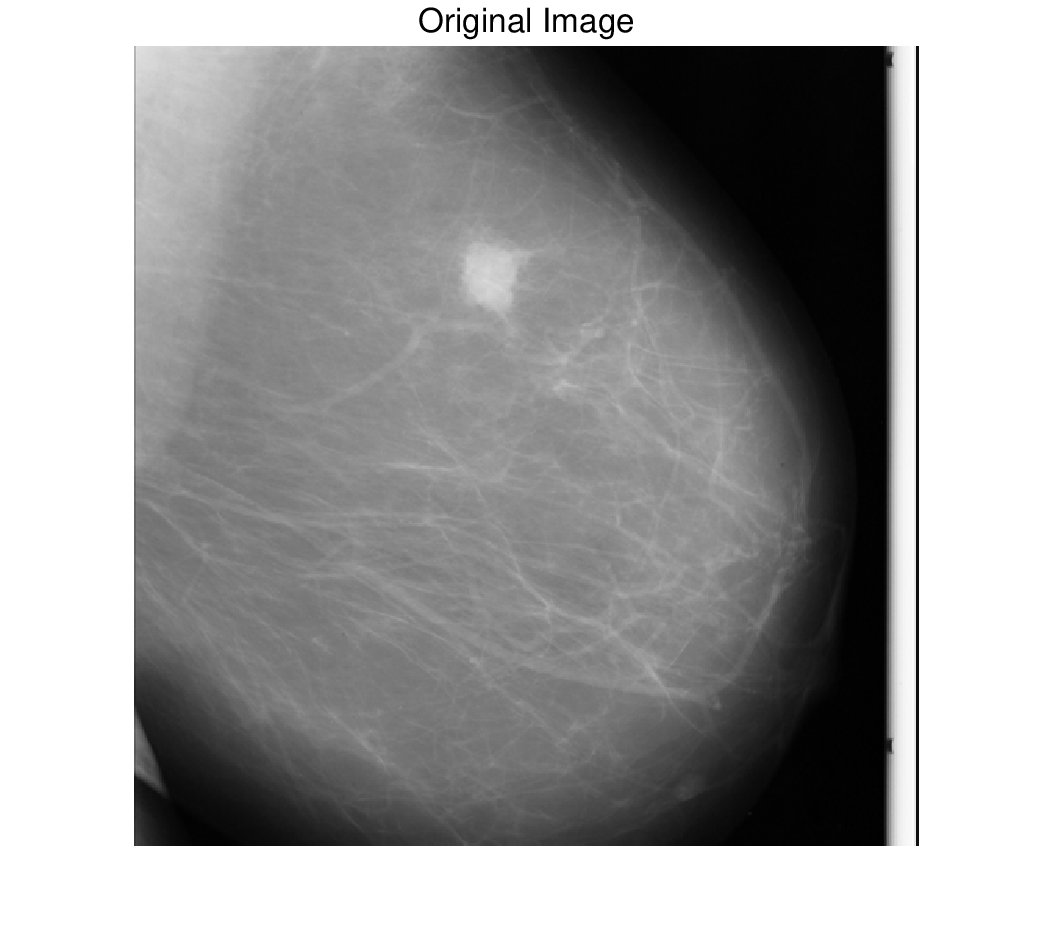}\\
		\vspace{0.02cm}
		\subcaption{}
		\includegraphics[width=0.75\linewidth,height=1\linewidth]{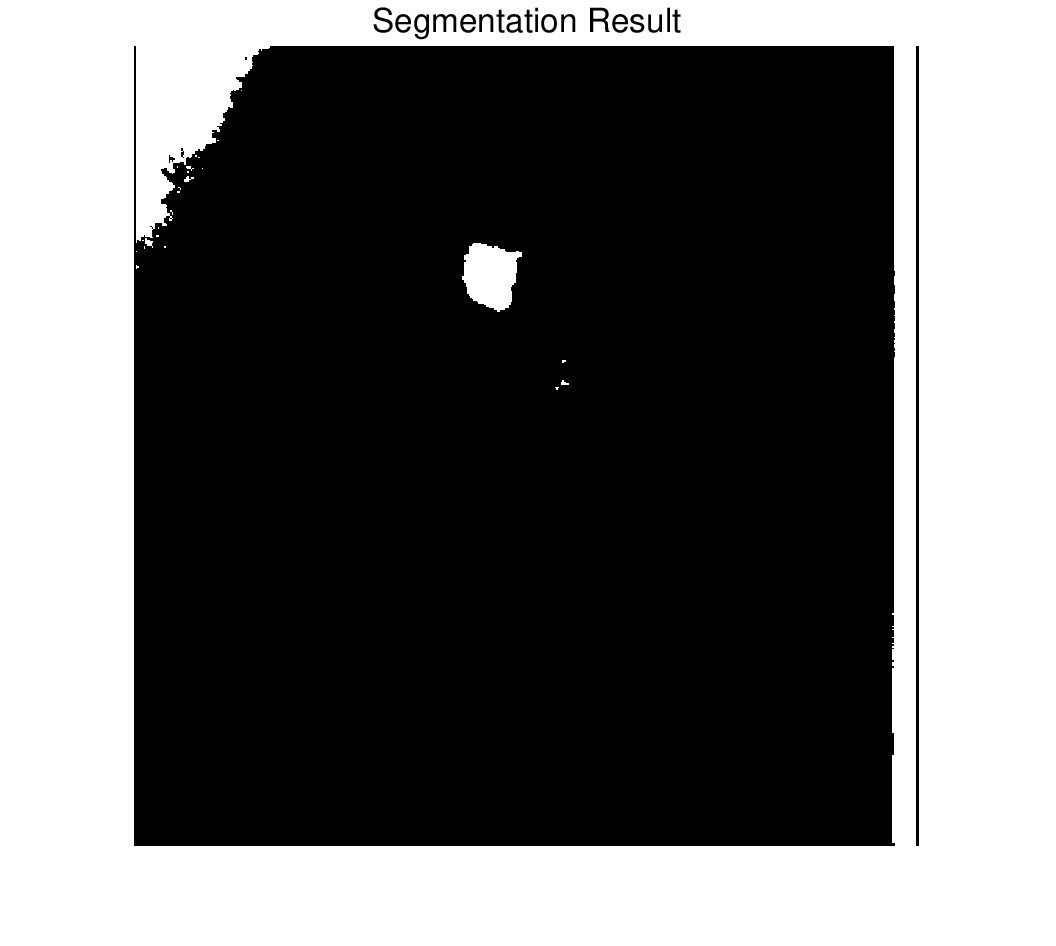}\\
		\vspace{0.02cm}
		\subcaption{}
	\end{minipage}
	\begin{minipage}[t]{0.15\linewidth}
		\centering
		\includegraphics[width=0.75\linewidth,height=1\linewidth]{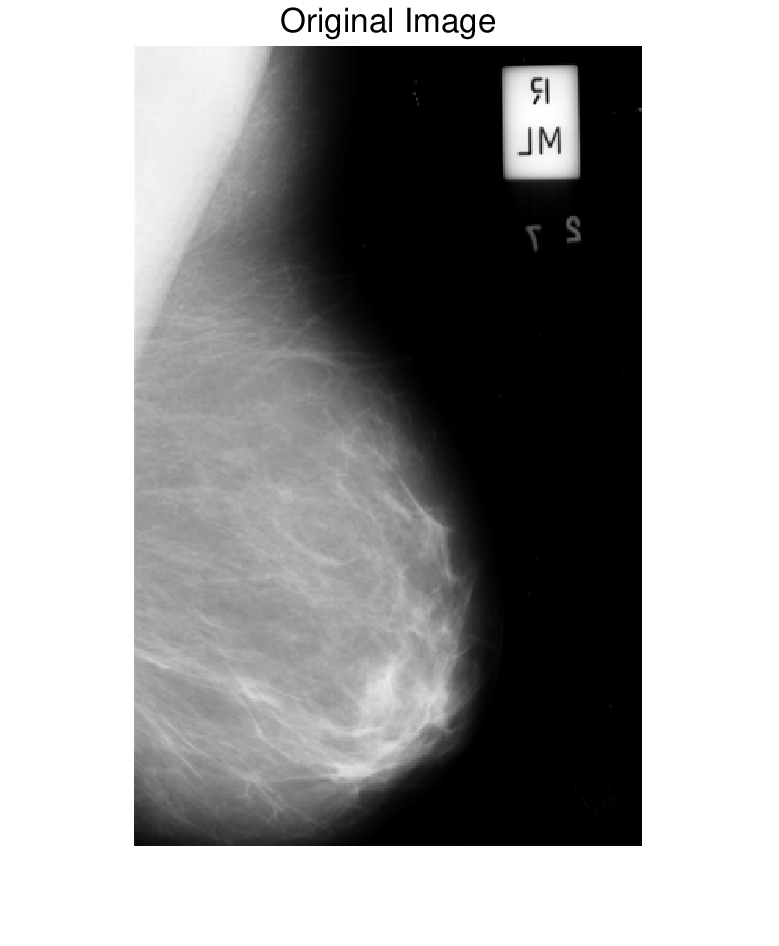}\\
		\vspace{0.02cm}
		\subcaption{}
		\includegraphics[width=0.75\linewidth,height=1\linewidth]{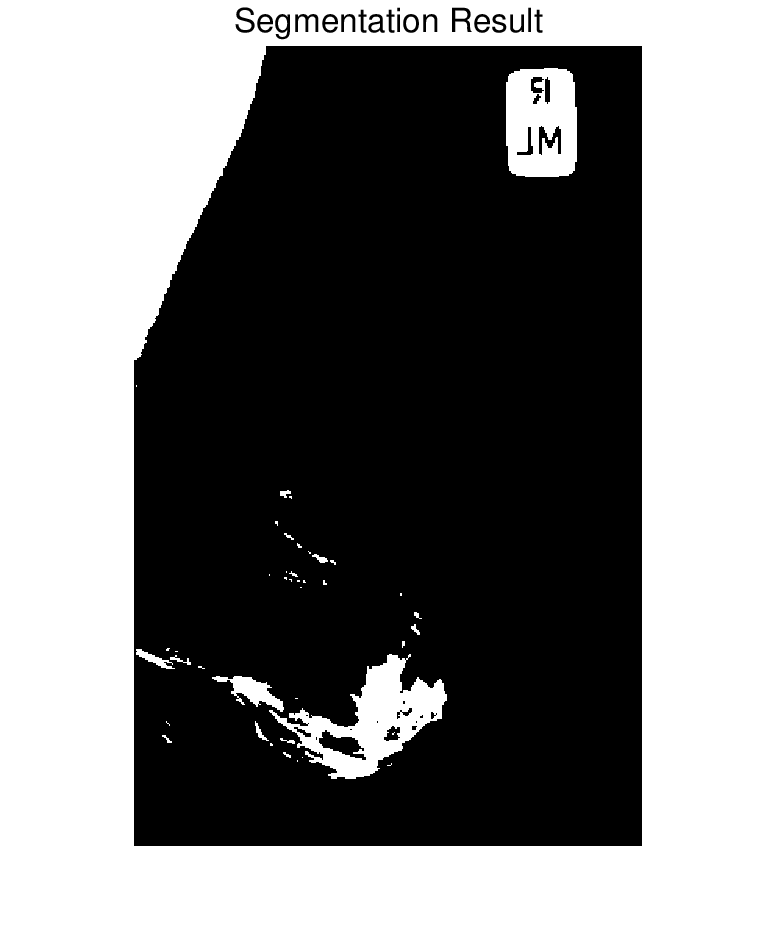}\\
		\vspace{0.02cm}
		\subcaption{}
	\end{minipage}%
	\begin{minipage}[t]{0.15\linewidth}
		\centering
		\includegraphics[width=0.75\linewidth,height=1\linewidth]{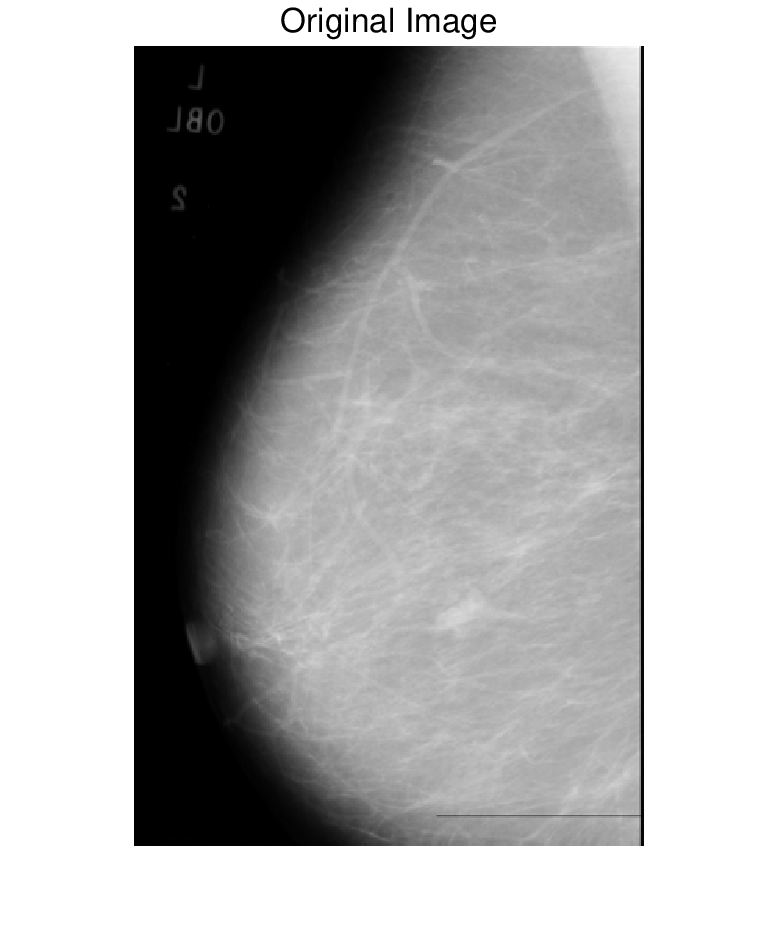}\\
		\vspace{0.02cm}
		\subcaption{}
		\includegraphics[width=0.75\linewidth,height=1\linewidth]{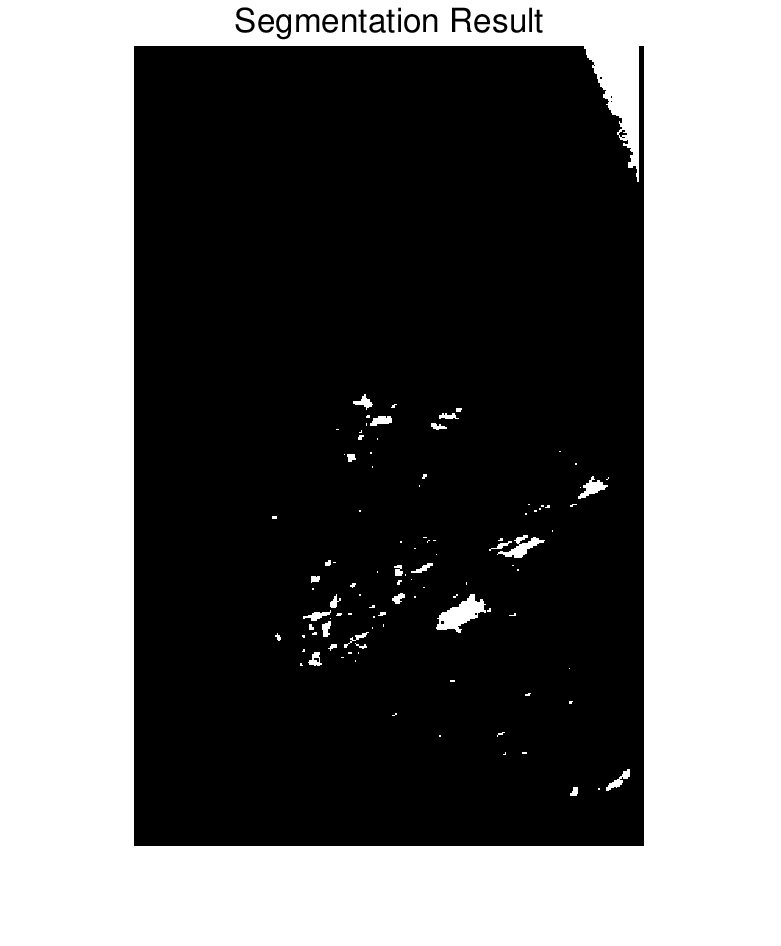}\\
		\vspace{0.02cm}
		\subcaption{}
	\end{minipage}%
	\begin{minipage}[t]{0.15\linewidth}
		\centering
		\includegraphics[width=0.75\linewidth,height=1\linewidth]{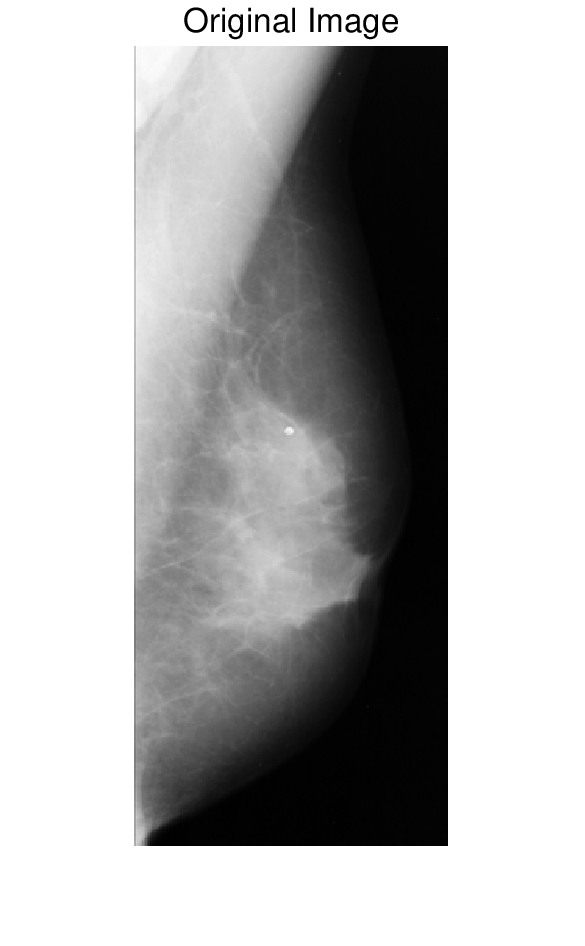}\\
		\vspace{0.02cm}
		\subcaption{}
		\includegraphics[width=0.75\linewidth,height=1\linewidth]{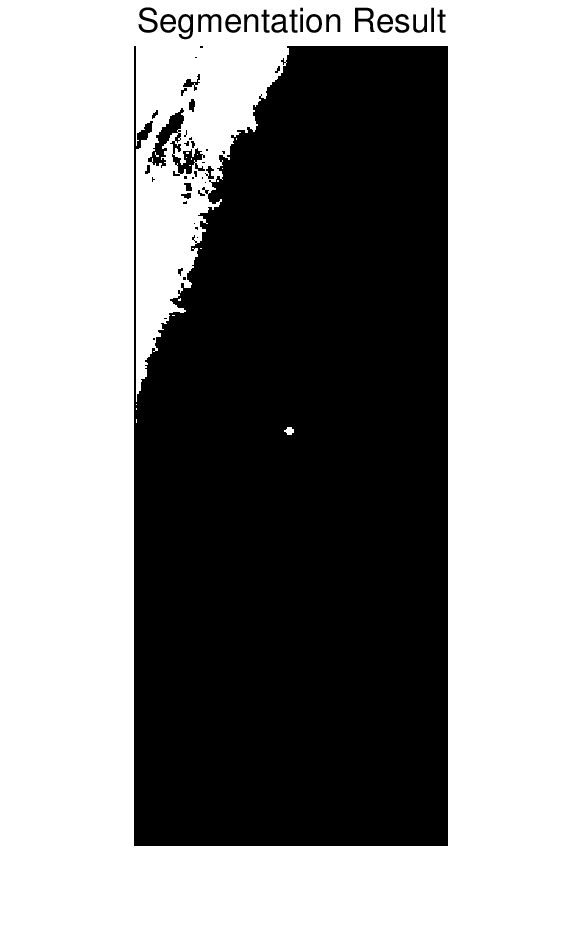}\\
		\vspace{0.02cm}
		\subcaption{}
	\end{minipage}%
	\centering
	\caption{Breast image segmentation results: (a) original image of mdb007; (b) segmentation results of mdb007; (c) original image of mdb030; (d) segmentation results of mdb030; (e) original image of mdb158; (f) segmentation results of mdb158; (g) original image of mdb120; (h) segmentation results of mdb120; (i) original image of mdb121; (j) segmentation results of mdb121; (k) original image of mdb238; (l) segmentation results of mdb238;}
	\label{fig:SR experiments}
\end{figure*}

The visual segmentation results are shown in Fig.\ref{fig:SR experiments}.
\par
Based on the segmentation results, our method can accurately distinguish breast masses from background tissue, as shown in Fig.\ref{fig:SR experiments}(e) and Fig.\ref{fig:SR experiments}(f). The images in \ref{fig:SR experiments}(g)-- (j) indicate that breast images of benign and malignant architectural distortion types have different segmentation results. Therefore, doctors can more easily distinguish benign and malignant architectural distortion. Combined with some morphological postprocessing tools, the algorithm can also predict the diagnostic result. From Fig.\ref{fig:SR experiments}(k) and Fig.\ref{fig:SR experiments}(l), our algorithm is effectively able to identify calcification points.
\par
\begin{figure}[htbp]
	\centering
	\begin{minipage}[t]{0.2\linewidth}
		\centering
		\includegraphics[width=0.75\linewidth,height=1\linewidth]{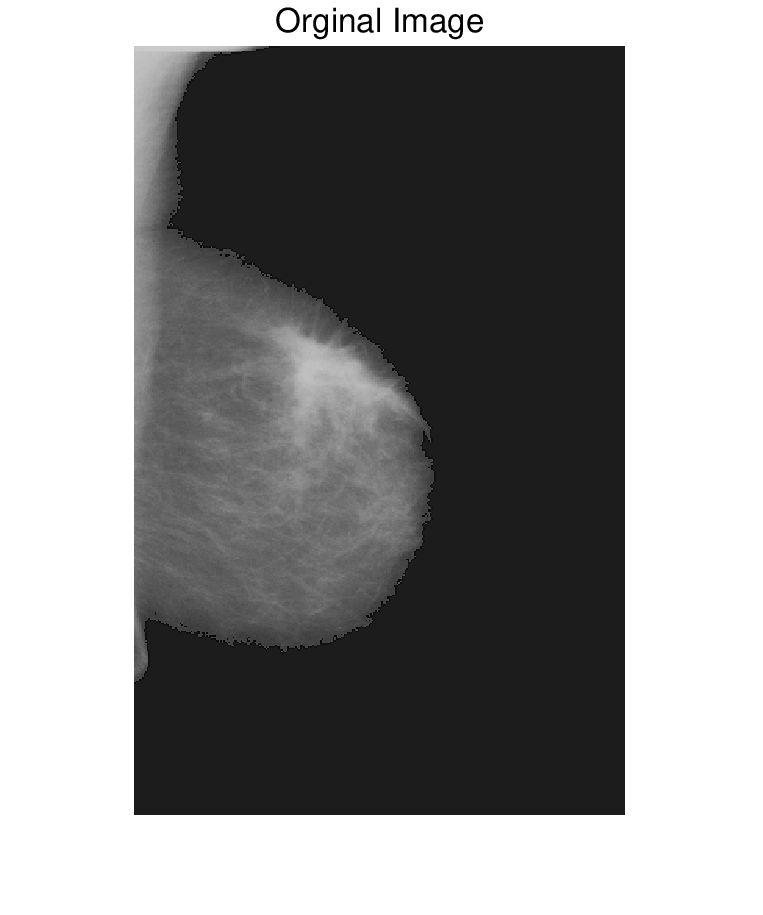}\\
		%\vspace{0.02cm}
		\subcaption{}
		\includegraphics[width=0.75\linewidth,height=1\linewidth]{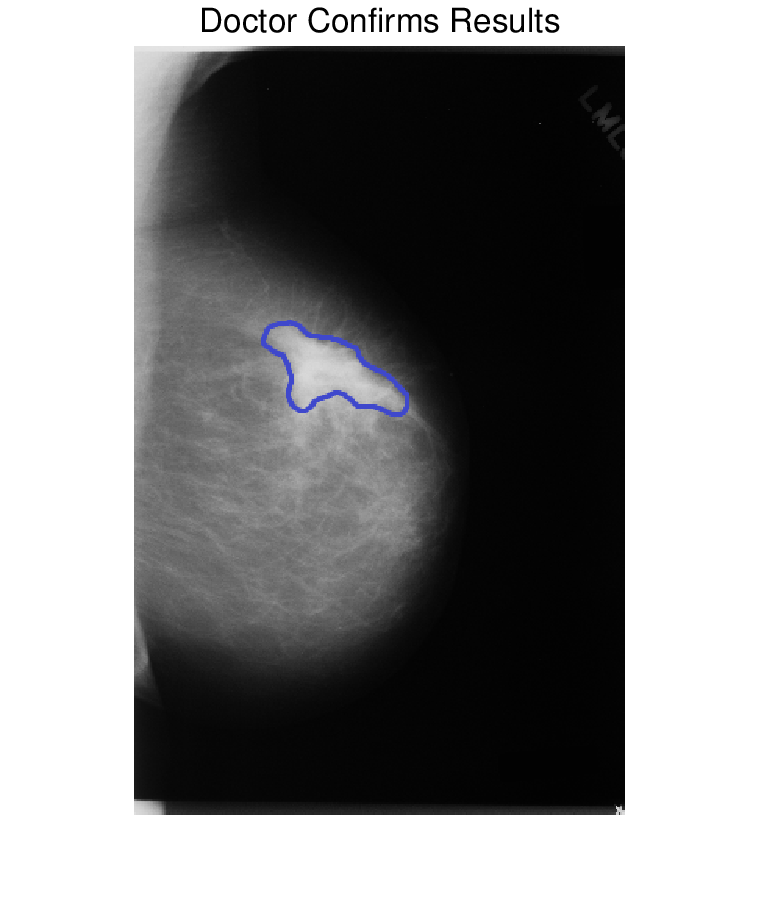}\\
		%\vspace{0.02cm}
		\subcaption{}
	\end{minipage}%
	\begin{minipage}[t]{0.2\linewidth}
		\centering
		\includegraphics[width=0.75\linewidth,height=1\linewidth]{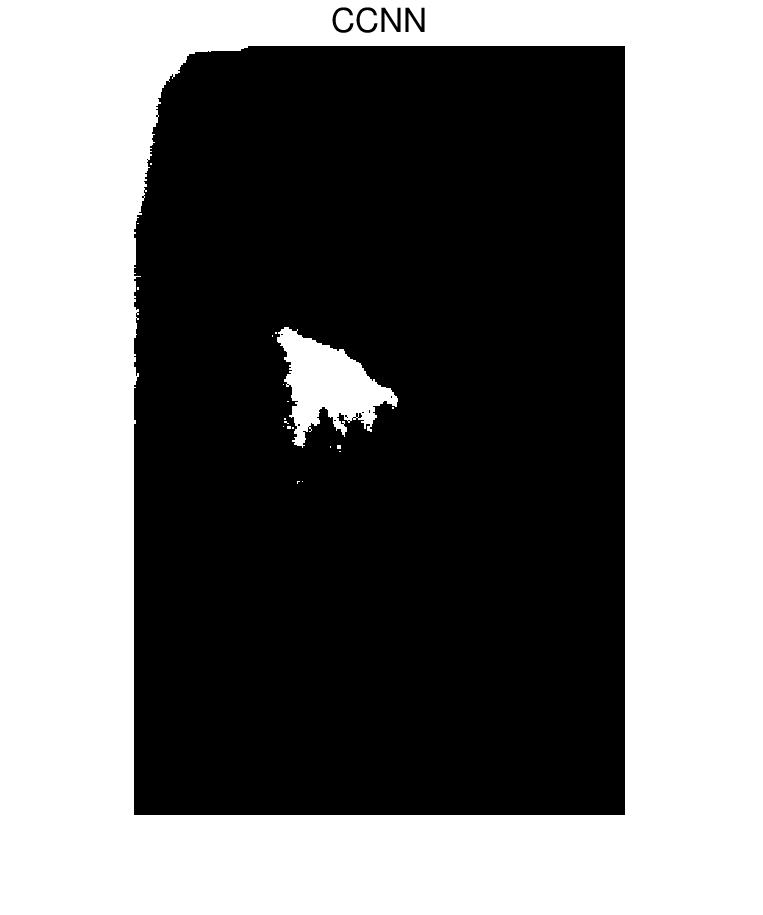}\\
		%\vspace{0.02cm}
		\subcaption{}
		\includegraphics[width=0.75\linewidth,height=1\linewidth]{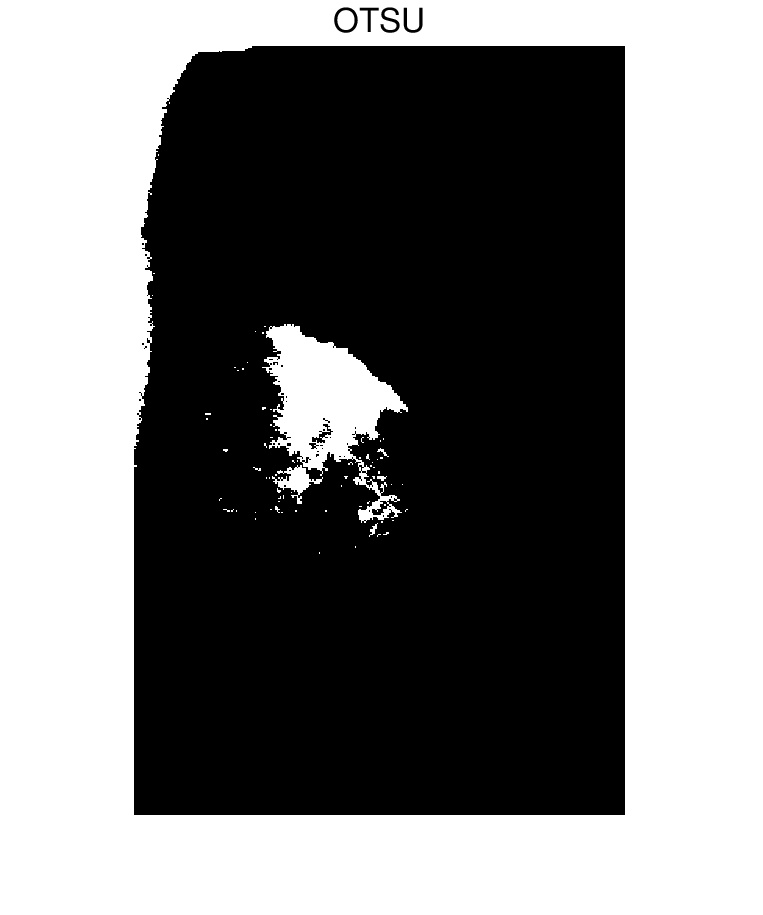}\\
		%\vspace{0.02cm}
		\subcaption{}
	\end{minipage}%
	\begin{minipage}[t]{0.2\linewidth}
		\centering
		\includegraphics[width=0.75\linewidth,height=1\linewidth]{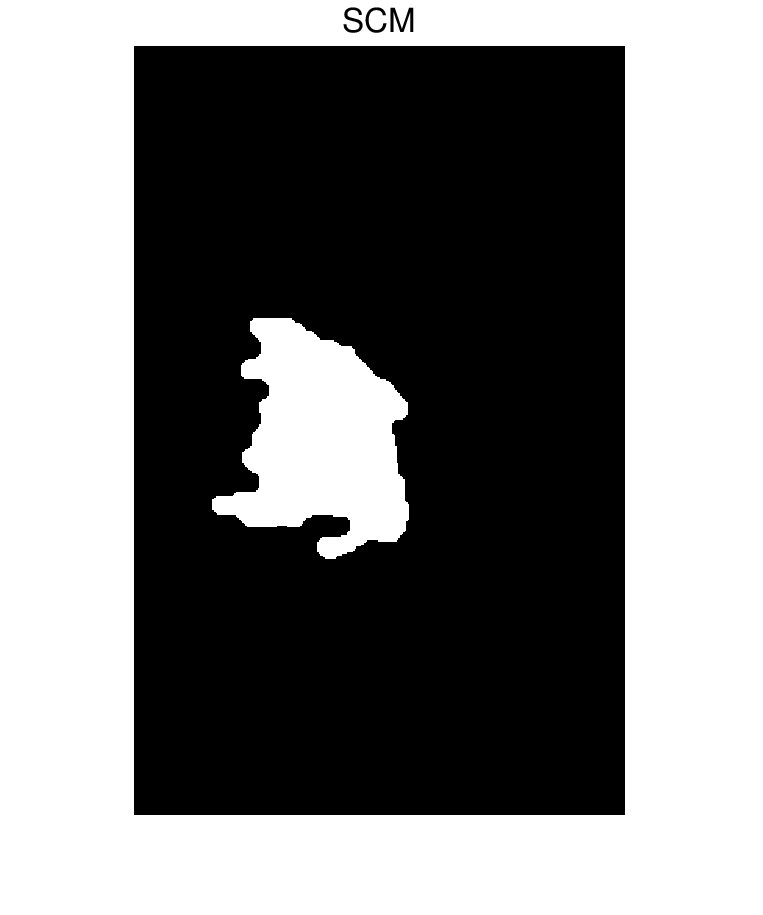}\\
		%\vspace{0.02cm}
		\subcaption{}
		\includegraphics[width=0.75\linewidth,height=1\linewidth]{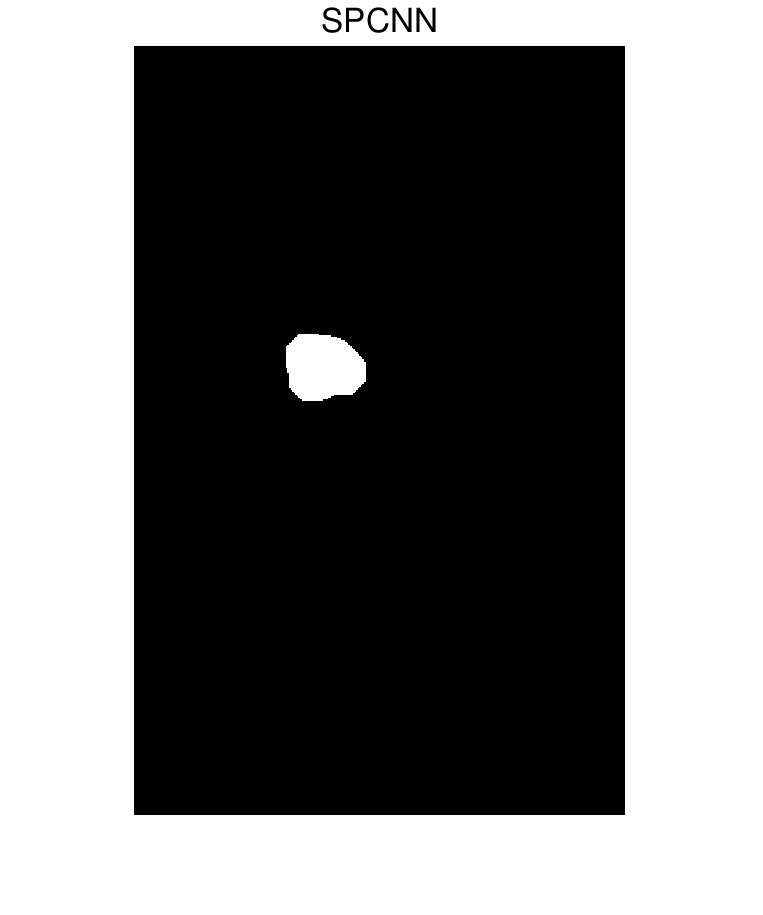}\\
		%\vspace{0.02cm}
		\subcaption{}
	\end{minipage}%
	\begin{minipage}[t]{0.2\linewidth}
		\centering
		\includegraphics[width=0.75\linewidth,height=1\linewidth]{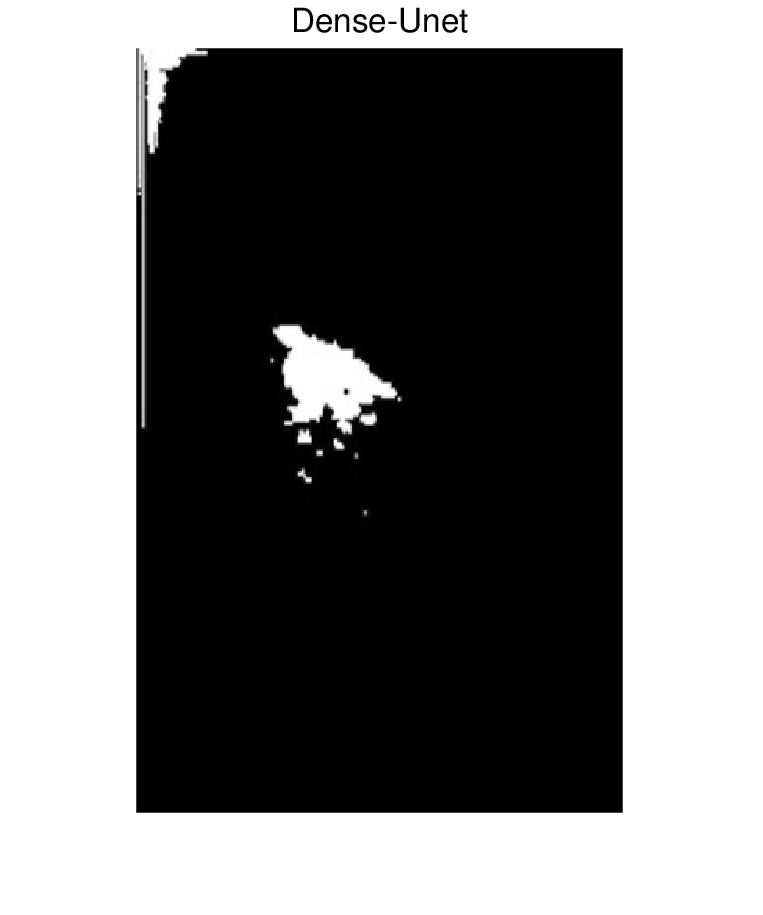}\\
		%	\vspace{0.02cm}
		\subcaption{}
		\includegraphics[width=0.75\linewidth,height=1\linewidth]{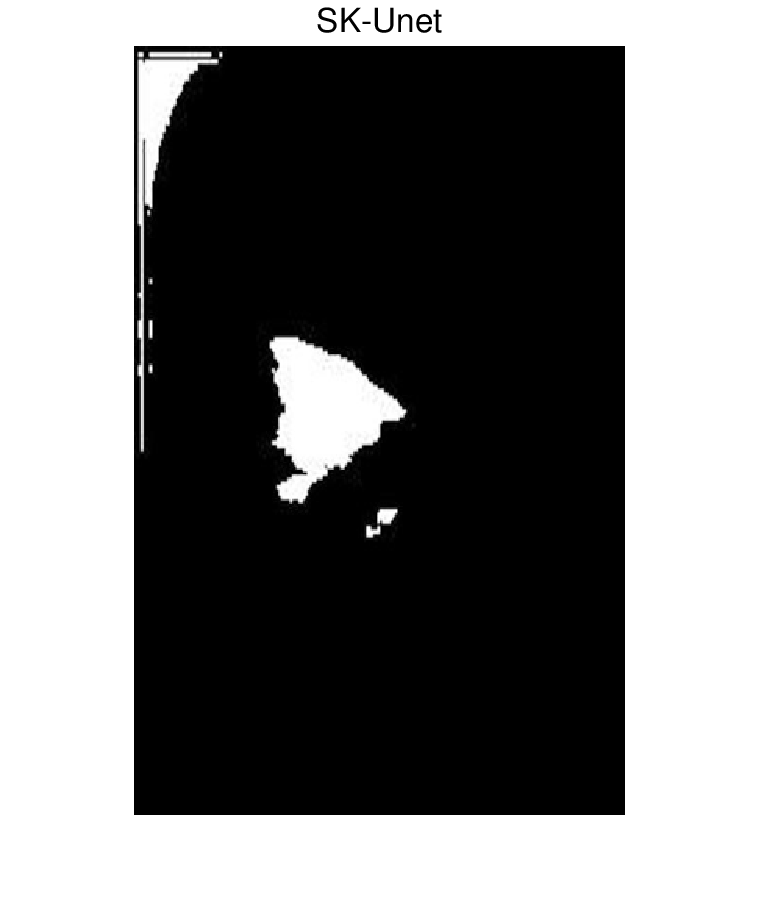}\\
		%	\vspace{0.02cm}
		\subcaption{}
	\end{minipage}%
	\centering
	\caption{Comparative breast image segmentation results: (a) original image from the DDSM database (well-defined/circumscribed masses); (b) segmentation results by a professional doctor; (c) segmentation results by CCNN; (d) segmentation results by OTSU; (e) segmentation results by SCM; (f) segmentation results by SPCNN; (g) segmentation results by Dense-Unet; (h) segmentation results by Sk-Unet;}
	\label{fig:SR comparison experiments}
\end{figure}

To compare our method with other segmentation methods, we randomly chose 100 mammogram images from the Digital Database for Screening Mammography (DDSM) as test images. A visual comparison of the segmentation results is shown in Fig. \ref{fig:SR comparison experiments}.
\par
Therefore, our method is able to accurately identify a single type of lesion without training, as shown in Fig.\ref{fig:SR experiments}(a)--Fig.\ref{fig:SR experiments}(f). However, it cannot identify lesions with well-defined/circumscribed masses and calcification, as shown in Fig.\ref{fig:SR experiments}(g)--Fig.\ref{fig:SR experiments}(h). In this situation, we believe that applying some morphological tools could obtain better results.
\par
We adopted Otsu, SPCNN, SCM, Dense-Unet and Selective Kernel U-Net as comparative algorithms to evaluate the effectiveness of our method. The Otsu method\cite{du2019improved} is an effective and adaptive thresholding scheme that has been widely applied for image segmentation. SCM and SPCNN are two state-of-art networks of visual cortex neural network model, which has superior performance in application of image segmentation\cite{zhan2015image,chen2011new}. Dense-Unet and Selective Kernel U-Net are deep learning method for biomedical image segmentation, which can provide good segmentation performance in medical image\cite{guan2019fully,byra2020breast}.
\par
We compared the algorithm segmentation results with the manual segmentation results of professional doctors and calculated the area overlap (OV) of the above algorithms. A visual comparison of the segmentation results is shown in Fig. \ref{tab:comparative results}. To quantitatively compare the above algorithms, we utilized five prevalent evaluation metrics, i.e., average pixel intensities (AVE), variance (VAR), area overlap (OV), sensitivity (SEN) and runtime (T). The results are shown in Tab.\ref{tab:comparative results}.
\begin{table}[htbp] 
	\centering
	\begin{tabular}{lccccl}
		\toprule
		Algorithms & AVE & VAR & OV(\%) & SEN & Time(s) \\
		\midrule
		Otsu & 0.4072 & 0.0103 & 0.0618 &\textbf{1} & 0.0894 \\
		SPCNN & 0.4073 & 0.0091 & 0.0666 &\textbf{1} & 0.0768 \\
		SCM & 0.5487 & 0.0140 & 0.2920 &0.9711 & \textbf{0.0686} \\
		CCNN & 0.9794 & 0.0058 & \textbf{0.8119} &0.6819 & 0.1540 \\
		Dense-Unet & \textbf{0.9990} & 0.0075 & 0.6849 & 0.9602 & 1.9470 \\
		SK-Unet & 0.9987 & \textbf{0.0184} & 0.7384 & 0.9225 &  7.8167\\
		\bottomrule
	\end{tabular}
	\caption{comparative results}
	\label{tab:comparative results}
\end{table}
\par
In this comparative experiment, the deep learning network did not achieve better results, which may be caused by the small sample size. Compared to all method, although our method requires a longer runtime than other unsupervised image segmentation method, its ability to accurately segment lesions is much better than that of the other methods. Therefore, the proposed segmentation method performs competitively compared to other prevalent algorithms.
\section{Conclusion and future work}
In this work, we proposed a novel model of the primary visual cortex. Compared with the previous best visual cortex neuron model, our CCNN model generates more types of spike trains under different stimuli and exhibits frequency doubling characteristics similar to those of real neurons. We compared the results of spike train data from CNN with those from real primary visual cortex neurons, and the results verified that our model is more similar to the real neurons.
\par
Because the results of existing medical imaging diagnoses rely on the subjective judgments of doctors and because the CCNN model is more similar to the visual cortex, it is natural to utilize our model to process medical images. In this work, we applied the CCNN model to a mammogram image segmentation task. The comparative results demonstrated that our method is superior at accurate lesion segmentation.
\par
The CCNN outputs different spike train patterns, which can be used as a type of "error" that can be backpropagated by algorithms, allowing CCNN to be used in deep learning networks.
\par
Thus, the advantage of our approach is that it further helps neurophysiologists understand how the primary visual cortex works and quantitatively predicts the temporal-spatial behavior of real neural networks. It also helps empower engineers to build more applications using visual neural networks and may inspire the next generation of deep learning networks for artificial intelligence purposes.
\par

In the future, some interesting work remains to be explored.
\par 
1. As is well known, visual neurons are sensitive to direction; however, it is unknown whether the CCNN model exhibits this behavior. If so, the mechanism of the neuron network should be explored.
\par
2. Brain waves are rhythmic oscillation patterns that can be registered as macroscopic oscillations utilizing EEG sensors on the scalp. Different rhythms are related to different brain activities: delta rhythmic are related to sleep states; theta might be entrance to further understanding learning and memory; alpha is usually related to attention, lucid thinking and integration; beta is present during the state of alert and problem solving; and gamma rhythms modulate perception and consciousness. It will be an interesting work to study whether our model can exhibit similar behavior as EEG signals. According to Baravalle et al, realized and imagined activities in the brain is different\cite{rosso2019characterization}. If our model can generate similar output sequence of EEG signal, perhaps we can distinguish between realized and imagined activities in the brain by studying the different sequences generated by the model. If so, the resulting works will help neurophysiologists further understand how the brain works. 
\par
3. To use the CCNN for artificial intelligence purposes, designing the ``error" that can be backpropagated by algorithms is an important task.
\par
4. The hardware circuits corresponding to the CCNN should be explored.
\par
5. The image segmentation algorithm proposed in the section V can only segment a single target object. The image segmentation algorithm of multi-target objects based on this model will be a meaningful work. 
\section*{Acknowledgments}
The data of Fig. \ref{fig:RealISI} were collected in the Laboratory of Dario Ringach at UCLA and downloaded from the CRCNS web site. https://crcns.org/data-sets/vc/pvc-1/conditions.
\par 
This study is supported by the Fundamental Research Funds for the Central Universities (No. 19lgpy230).
\par
Conflict of Interest: The authors declare that they have no conflicts of interest.

\bibliography{refs}

\end{document}